%% file: main.tex
\documentclass[12pt]{article}
\usepackage{scicite}

\usepackage{times}
\usepackage{comment}
\usepackage{setspace}
\usepackage{lineno}

\usepackage{amsmath,amsfonts,amssymb}   
\usepackage{graphicx}  
\usepackage{float}     
\usepackage{graphicx} 
\usepackage{bm}
\usepackage{textcomp}
\usepackage{physics}
\usepackage{mathtools}
\usepackage{tocloft}
\usepackage[labelfont=bf]{caption}
\usepackage[margin=.9in]{geometry}
\usepackage{wrapfig}
\usepackage[normalem]{ulem}

\usepackage[margin=.9in]{geometry}
\usepackage[labelfont=bf]{caption}
\usepackage[font=footnotesize]{caption}
\usepackage{xcolor}
\usepackage{soul}

\newcommand{\textib}[1]{\textit{\textbf{#1}}}
\newcommand{\yl}[1]{\color{black} {#1} \color{black}}

\usepackage{filecontents}
\usepackage{multibib}
\newcites{supp}{Methods References}
\newcites{suppl}{References and Notes}

\newenvironment{sciabstract}{%
\begin{quote} \bf}
{\end{quote}}

\newcounter{lastnote}

\title{\yl{0ptical trapping with optical magnetic field and photonic Hall effect forces}}

\author
{
Yanzeng Li$^{1,*}$, Emmanuel Valenton$^{1,2}$, Spoorthi Nagasamudram$^{1,3}$,\\
John Parker$^{4}$, Marcos Perez$^{5}$, Uttam Manna$^{5}$, Mahua Biswas$^{5}$,\\
Stuart A. Rice$^{1,2}$, Norbert F. Scherer$^{1,2,*}$\\
\\
\normalsize{$^{1}$James Franck Institute, The University of Chicago, Chicago, IL 60637, USA}\\
\normalsize{$^{2}$Department of Chemistry, The University of Chicago, Chicago, IL 60637, USA}\\
\normalsize{$^{3}$Department of Physics, The University of Chicago, Chicago, IL 60637, USA}\\
\normalsize{$^{4}$Department of Physics, The University of Chicago, Chicago, IL 60637, USA}\\
\normalsize{$^{5}$Department of Physics, Illinois State University, Normal, IL 61790, USA}\\
\\
\normalsize{$^\ast$To whom correspondence should be addressed;}\\
\normalsize{E-mail:  nfschere@uchicago.edu; yanzengli@uchicago.edu.}
}

\date{}

\usepackage{xr}
\makeatletter

\newcommand*{\addFileDependency}[1]{
\typeout{(#1)}
\@addtofilelist{#1}
\IfFileExists{#1}{}{\typeout{No file #1.}}
}\makeatother

\newcommand*{\myexternaldocument}[1]{%
\externaldocument{#1}%
\addFileDependency{#1.tex}%
\addFileDependency{#1.aux}%
}

\myexternaldocument{SI}
\begin{document}

\maketitle

\begin{sciabstract}
\yl{Optical trapping is having ever-increasing impact in science --- particularly biophysics\cite{ashkin1987optical,bustamante2021optical,fazal2015real}, photonics\cite{urban2010laser,do2013photonic,bao2014optical,violi2016light} and most recently in quantum optomechanics\cite{delic2020cooling,magrini2021real,tebbenjohanns2021quantum} --- owing to its superior capability for manipulating nanoscale structures and materials. However, essentially all experimental optical trapping studies in the optical dipole regime have, to date, been dominated by the interaction between a material's electric polarizability, \textbf{\(\alpha_\text{e}\)}, and the electric part of the incident electromagnetic field, and therefore described by electric field intensity gradient forces\cite{novotny2012principles,grier2003revolution}. Optical trapping based on \textit{optical magnetic} light-matter interactions has not been experimentally addressed despite it's immediate extension of the boundaries of optical trapping research and applications. This paper addresses this long-standing deficiency through the realization of \textit{optical magnetic trapping} of large index of refraction (i.e., Si) nanoparticles and also presents a formalism for quantitative understanding of the experimental findings. Our experimental optical trapping results require including optical magnetic polarizability, \textbf{\(\alpha_\text{m}\)}, and electric-magnetic scattering forces associated with the Photonic Hall effect\cite{hosten2008observation} that are qualitatively and quantitatively validated by Maxwell stress tensor calculations. Our findings bring new opportunities for nanoparticle manipulation, potentially relax the limitations Ashkin claimed based on the optical Earnshaw's theorem\cite{Ashkin1983stability}, motivate optical matter formation\cite{dholakia2010colloquium,burns1989optical,burns1990optical,ng2005photonic,wang2016optically,han2018crossover} by optical magnetic interactions, and suggest new N-body effects\cite{parker2020optical,nan2022creating} and symmetry breaking\cite{yifat2018reactive} to drive dynamics of optical matter systems.}
\end{sciabstract}

Since its inception in 1970\cite{Ashkin1970}, optical trapping has developed into a large field of investigation and application, catalyzing substantial advances in diverse disciplines such as biophysics\cite{ashkin1987optical,bustamante2021optical,fazal2015real}, material science\cite{urban2010laser,do2013photonic,bao2014optical,violi2016light}, and macroscopic quantum physics\cite{delic2020cooling,magrini2021real,tebbenjohanns2021quantum}. Even after five decades of research, new topics in optical trapping continue to emerge. These include fascinating physical phenomena such as optical matter formation\cite{dholakia2010colloquium,burns1989optical,burns1990optical,ng2005photonic,wang2016optically,han2018crossover}, many-body photonic interactions and symmetry breaking-induced dynamics leading to optical matter machines\cite{parker2020optical, nan2022creating},  symmetry-breaking electrodynamics\cite{dogariu2013optically,ivlev2015statistical,chvatal2015binding,sukhov2015actio,karasek2017dynamics,simpson2017optical,sule2017rotation,yifat2018reactive,peterson2019controlling}, and mesoscopic dipole interactions\cite{arita2018optical,svak2021stochastic,rieser2022tunable}.

All of the aforementioned studies and essentially all of optical trapping heretofore utilize the optical force associated with the electric fields and gradients\cite{novotny2012principles,grier2003revolution},
\begin{equation}
	\label{eqn:dpfe}
    \langle\textib{F}_{\text{e}}^{\text{PD}}\rangle=\frac{\alpha^{'}_{\text{e}}}{4}\nabla|\textib{E}|^2+\frac{\alpha^{''}_{\text{e}}}{2}|\textib{E}|^2\nabla\Phi
\end{equation}
\noindent \yl{while neglecting the (possible) magnetic counterpart (see the Methods for a more general form for Eq.~\ref{eqn:dpfe}).} The first and second terms in Eq.~\ref{eqn:dpfe} represent: (i) the intensity gradient force associated with the incoming electric field $E$ and real part of the electric polarizability $\alpha_{\text{e}}^{'}$; and (ii) the radiation pressure associated with the gradient phase $\Phi$ of the electric field (i.e., the local curvature of the wavefront) and the imaginary part of the electric polarizability $\alpha_{\text{e}}^{''}$. \yl{Although optical beams are electro-magnetic fields, where both their electric and magnetic components can potentially interact with matter, universal adoption of Eq.~\ref{eqn:dpfe} seems reasonable since induced magnetizations are practically zero at optical frequencies\cite{novotny2012principles}.} 

\yl{Still, if one considers the possibility of optical magnetic forces, the most simple expectation is based on a point-dipole model\cite{nieto2010optical} that the magnetic field-mediated force behaves analogous to the electric version. That is,
\begin{equation}
    \label{eqn:dpfm}
    \langle\textib{F}_{\text{m}}^{\text{PD}}\rangle=\frac{\alpha'_{\text{m}}}{4}\nabla|\textib{H}|^2+\frac{\alpha''_{\text{m}}}{2}|\textib{H}|^2\nabla\Phi,
\end{equation}
\noindent where all terms are the magnetic (field) analogs of those in Eq.~\ref{eqn:dpfe}. This expression is analogous to what is given in Ref\cite{zeng2022direct}; see the Methods for a generalized form of Eq.~\ref{eqn:dpfm}. The magnetic field-associated gradient force (first term) says that positive optical magnetization would give rise to an attractive force. Yet, despite some recent theoretical work\cite{nieto2010optical,xu2020kerker} suggesting possible routes for experimental confirmation using magnetodielectric nanoparticles with their inducible magnetizations, experimental evidence of optical magnetic forces being important in optical trapping is still lacking. The exception is a recent experimental study of photoinduced magnetic dipolar forces using (atomic) force microscopy\cite{zeng2022direct}. }

\yl{There are also unresolved fundamental questions concerning induced optical magnetic forces.} In particular, Novotny predicted\cite{novotny2012principles} that induced magnetization in optical trapping would behave opposite to what induced electric polarizations do; i.e., a particle with a \underline{negative} induced magnetization will be attracted to a magnetic field maximum. \yl{This suggestion contradicts the expectation based on Eq.~\ref{eqn:dpfm}.} These conflicting perspectives urgently demand confirmation of the underlying mechanism of \yl{optical magnetic field trapping} specifically and light-matter interactions generally.

\yl{Although Eqs.~\ref{eqn:dpfe} and \ref{eqn:dpfm} are suitable for point particles (and as we show later for off-resonance conditions), a correct description of matter-radiation interactions for finite size particles involves higher-order electric and magnetic multipole modes (and cross-terms), resulting in complicated theoretical expressions that complicate interpretation of experiments. However, if the wavelength of the optical trapping beam is on-resonance, these modes will dominate the interaction. In the present study, the electric dipole (ED) and magnetic dipole (MD) modes are on-resonance and Eqs. \ref{eqn:dpfe}-\ref{eqn:tsf} suffice for interpretation of our experimental findings. }

Here, we experimentally demonstrate \textit{optical magnetic field trapping} of large refractive index nanoparticles and provide a new quantitatively accurate theoretical description thereof. \yl{Large index of refraction Silicon (Si) nanoparticles of various sizes were investigated exploring the spectral detuning of their Mie optical magnetic dipole resonance relative to the fixed trapping laser wavelength.} We observed that when optically excited at their magnetic dipole resonance, single Si nanoparticles selectively (thermally) sample the most intense magnetic field regions of focused vector beams\cite{youngworth2000focusing}. However, application of the point dipole optical magnetic (magnetization) model of Eq.~\ref{eqn:dpfm} cannot explain all of our observations. In particular, when the Si nanoparticle diameter is large, it has \textit{negative optical magnetic polarizability} and by Eq.~\ref{eqn:dpfm} should be repelled from regions of strong optical magnetic fields, but it is not. \yl{We show that the full range of optical trapping observations can be explained by including forces arising from the \textit{Photonic Hall Effect} (PHE)\cite{onoda2004hall, hosten2008observation, haefner2009spin, rodriguez2010optical} that is demonstrated with more details in Sect.~\ref{SI_subsect:PHE} in the Supplementary Information.} We develop a scattering-corrected model (SCM) using generalized Lorentz-Mie theory (GLMT)\cite{maheu1988concise,gouesbet1988computations,lock2009generalized} that includes the PHE-associated optical force, which is a nonconservative scattering force that manifests an anisotropy associated with the polarization of the trapping laser field. \yl{The optical magnetism and SCM-associated forces are validated by Maxwell stress tensor (MST) calculations. Extension to other high index material-based nanoparticles is demonstrated (see Sect.~\ref{SI_subsect:variousMaterials} in the Supplementary Information).}

\begin{figure}[h!]
	\centering
	\hspace*{-0.5cm}
	\includegraphics[scale=.5]{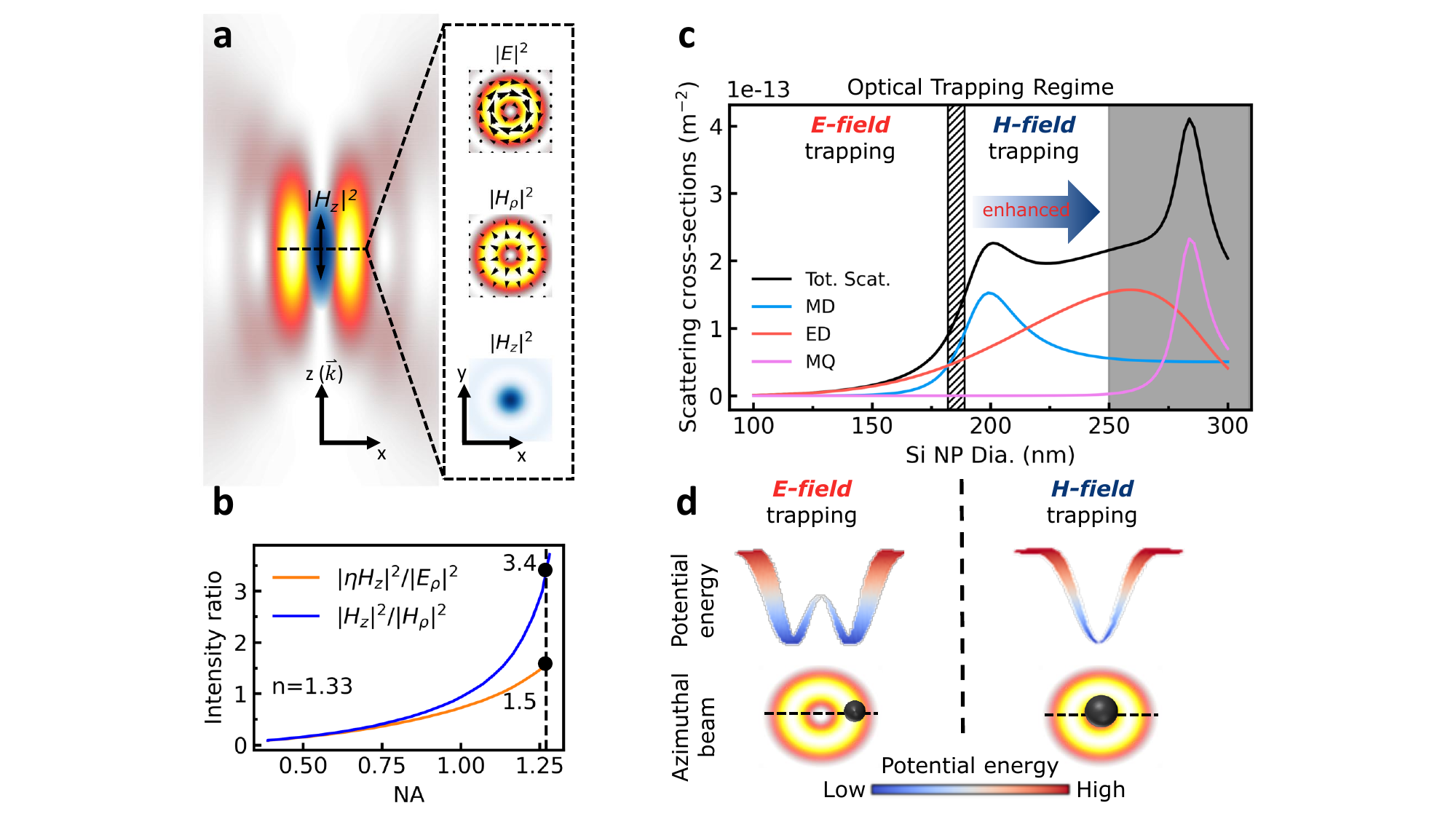}
	\caption{\textbf{Optical magnetic trapping of Si nanoparticles with an azimuthally polarized beam.} 
		\textbf{(a)} The intensity profile of a tightly focused azimuthally polarized beam decomposed into three constituent fields in the focal volume: $E$ (reddish), $H_{\rho}$ (reddish), and $H_{z}$ (blue), with black arrows representing their polarization directions.
		\textbf{(b)} The ratios of the $H_{z}$ to $E$ and $H_{\rho}$ intensities as a function of numerical aperture (NA) of the focusing lens. The black dots in the plot indicate the ratios for an NA of 1.27, which corresponds to the microscope objective used for trapping Si nanoparticles in water (n=1.33). 
		\textbf{(c)} The scattering cross sections of Mie multipole resonances of Si nanoparticles are calculated at the 770~nm trapping laser wavelength. The first intersection between the MD and ED scattering cross-section curves occurs at Si nanoparticle dia.=182~nm. This nanoparticle size separates the optical trapping into E-field- and H-field-dominated trapping domains. The hatched and grey regions represent the E-H transition (182-190 nm) and higher order multipole dominated  ($>$250 nm) regions, respectively. Si nanoparticles with sizes from about 180-250 nm in dia. will manifest trapping forces with strong optical magnetic contributions.
		\textbf{(d)} The calculated potential energy profiles across the focused azimuthally polarized vector beam illustrate the expected (energetic) tendencies of Si nanoparticles to trap in the E- and H-field dominated portions of the optical beam.
	}
	\label{fig:fig1}
\end{figure}

\section*{Results}
\subsection*{Optical beams and materials with strong optical magnetic properties}
Our operational hypothesis for demonstrating optical magnetic field trapping requires: (i) a focused electromagnetic beam that allows distinguishing the electric and magnetic fields; and (ii) nanoscale size materials that (strongly) interact with optical magnetic fields.
Fig.~\ref{fig:fig1}a shows the result of an electrodynamics simulation of a focused azimuthally polarized vector beam\cite{youngworth2000focusing}. The segregated magnetic field is primarily a longitudinally polarized component, $H_z$, at the beam center (where the electric field vanishes) together with the azimuthally polarized electric component, $E$\cite{youngworth2000focusing,novotny2012principles,manna2017selective}. \yl{There is an additional magnetic component of light with radial polarization, $H_{\rho}$, that shares the annular region with the electric component.} Figure~\ref{fig:fig1}b shows that the intensity of the longitudinally polarized magnetic field, $H_z$, increases with the tightness of the beam focus; the ratios $|\eta H_z|^2$/$|E|^2$ and $|H_z|^2$/$|H_{\rho}|^2$ increase with the numerical aperture (NA) of the objective (shown from NA~=~0.4 to 1.29)\cite{youngworth2000focusing}. When the azimuthally polarized beam is tightly focused with a NA of 1.27 (our experimental condition) the longitudinal magnetic field $H_z$ attains an intensity 1.5 and 3.4 times greater than the transverse electric $E$ and magnetic $H_{\rho}$ field intensities, respectively.
The calculation assumes a water environment (n=1.33) and considers the wave impedance $\eta=\sqrt{\epsilon_w/\mu_w}$ for the magnetic-to-electric intensity ratio where $\epsilon_w$ and $\mu_w$ are water's permittivity and permeability.
This focused trapping beam would enhance the $H_z$ associated optical force and reduce any competing forces due to other fields at the beam center thereby satisfying one condition for selective optical magnetic trapping. 

Our demonstration of optical magnetic trapping also requires a material with a strong \textit{optical magnetic dipole resonance}. Si nanoparticles satisfy this condition\cite{kuznetsov2012magnetic} (see Fig.~\ref{fig:fig2}d) and are small enough that they can interact dominantly with the longitudinally polarized magnetic field $H_z$ via their Mie magnetic dipole (MD) resonance. This electromagnetic field-nanoparticle interaction is expected to create optical magnetic forces as described by Eq.~\ref{eqn:dpfm}. However, as suggested in Fig.~\ref{fig:fig1}c and shown explicitly in Fig.~\ref{fig:fig2}d, Si nanoparticles also have a strong (and spectrally broad) electric dipole (ED) resonance\cite{kuznetsov2012magnetic} that can be excited by the annularly distributed electric field $E$. Therefore, electric field-mediated forces (Eq.~\ref{eqn:dpfe}) can also act on the Si nanoparticles. \yl{The magnitude of these two types of forces depend not only on the intensity of the respective fields, but also on the strength of the field-particle interactions that follow from the single Si nanoparticle ED and MD scattering cross-sections at the trapping laser wavelength (see Fig.~\ref{fig:fig2}d).} Figure~\ref{fig:fig1}c presents a GLMT calculated scattering spectrum (in black) of Si nanoparticles as a function of diameter for the 770~nm trapping laser wavelength. The total scattering is decomposed into the MD (blue) and ED (red) mode contributions. The MD mode dominates over the ED mode only in a narrow size range centered at a diameter of 200~nm, which corresponds to the peak of the MD resonance (for 770~nm trapping laser wavelength). This diameter-dependent scattering suggests another operational hypothesis that only Si nanoparticles in a certain size range (as shown in Fig.~\ref{fig:fig1}c) will manifest optical magnetic field trapping.

\subsection*{Optical trapping of Si nanoparticles}
Testing the aforementioned hypothesis and verifying optical magnetic trapping is accomplished by observing where the Si nanoparticles are most probable to be localized in the focused azimuthally polarized trapping beam. \yl{Since the experiments are conducted in room temperature solution, the random Brownian forces enable statistical (thermal) sampling of force interactions in the inhomogeneous trapping field (Fig.~\ref{fig:fig1}a).} The potential energy profiles of Fig.~\ref{fig:fig1}d for Si nanoparticles in the beam show that the energy minimum changes from the azimuthally polarized beam's annular E-field region to its center when the dominant electromagnetic interaction (and trapping force) changes from ED mode dominated to MD mode dominated with increasing Si nanoparticle size. Note that the upper size boundary that we consider in the experiments is when the magnetic quadruple (MQ) mode in Fig.~\ref{fig:fig1}c begins to dominate the scattering (at dia.$>$250~nm). Optical forces due to higher-order modes become pronounced (eventually dominant) beyond this Si nanoparticle size (see Fig.~\ref{fig:high-order_force}) and fall outside the scope of this study. Furthermore, we do not consider thermophoresis\cite{duhr2006molecules} due to the small absorbance of Si nanoparticles at 770~nm (see Fig.~\ref{fig:SiNP_apt_spectrum}).

\begin{figure}[t!]
	\centering
	\hspace*{-0.5cm}
	\includegraphics[clip,trim=0cm 4cm 0cm 3.7cm,scale=.5]{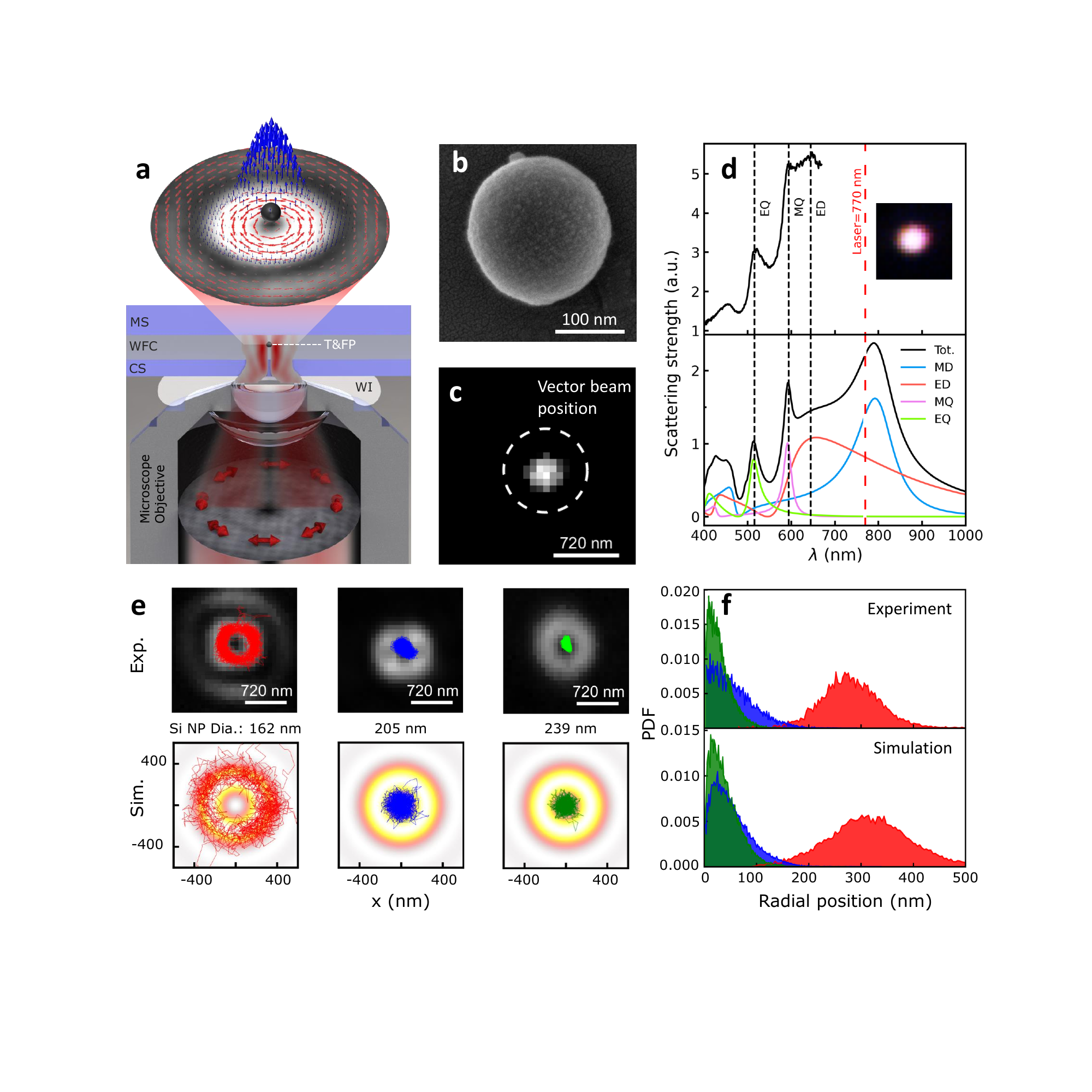}
	\caption{\textbf{Experimental evidence of optical magnetic trapping of Si nanoparticles.} 
		\textbf{(a)} Schematic representation of the microscope objective that is employed to generate a tightly focused azimuthally polarized beam for trapping Si nanoparticles in a water chamber. The closeup displays a Si nanoparticle at the trapping plane of the azimuthally polarized beam, with arrows indicating the longitudinal and azimuthal polarization of $H_z$ (blue) and $E$ (red), respectively.
		\textbf{(b)} SEM micrograph of a drop-cast Si nanoparticle that coincidentally has the same dimensions as the trapped one shown in panel \textbf{(c)}.
            \textbf{(c)} Dark-field microscopy image of a representative Si nanoparticle with a diameter of 205 nm trapped at the center of the focused azimuthally polarized beam delineated by the white dashed circle.
		\yl{\textbf{(d)} The size of the Si nanoparticle shown in panel \textbf{(c)} is determined by comparing the particle's \textit{in-situ} measured scattering spectrum (upper panel) with calculated Mie scattering spectra (lower panel; calculated in a water environment) and adjusting the Si nanoparticle size in simulation until their scattering spectral signatures align, including the multi-pole modes. The red dashed vertical line indicates the trapping laser wavelength.}
		\textbf{(e)} Nanoscale localizations of three representative Si nanoparticles with different sizes overlaid with their measured trapping beam profiles as gray-scale background. The upper and lower rows exhibit the experimental and GLMT-Langevin dynamics simulation results, respectively. 
  \textbf{(f)} The associated experimental (upper) and simulation (lower) radial position distributions.
	}
	\label{fig:fig2}
\end{figure}

Figure~\ref{fig:fig2}a presents a schematic of optical trapping where a water-immersion microscope objective (Nikon CFI SR Plan Apo IR 60X) focuses an azimuthally polarized beam into a thin (120~\(\mu\)m) water-filled chamber with colloidal Si nanoparticles. The expanded view depicts the azimuthally polarized beam focused near the top water-glass interface, with arrows indicating the (local) polarization directions and spatial distributions of both the longitudinal magnetic $H_z$ (blue) and electric $E$ (red) fields. The Si nanoparticle located at the maximum of the longitudinal magnetic field (i.e., at the center of the beam) provides a visual representation of magnetic field trapping. Details about Si nanoparticle synthesis and experimental implementation are given in the Supplementary Information. 

Scanning electron microscopy (SEM) was employed for \textit{ex-situ} characterization of particle size and shape. Figure~\ref{fig:fig2}b shows a SEM micrograph of a representative Si nanoparticle that was prepared on a substrate using a standard drop-casting technique. The optical scattering spectrum of this Si nanoparticle (not shown) is very similar to the one shown in Fig.~\ref{fig:fig2}d. The SEM measurement confirms that the comparison of experimental and calculated optical scattering spectra give an accurate determination of nanoparticle size with tolerance of $\pm$2~nm. Furthermore, well-resolved peaks in the measured scattering spectra indicate spherical nanoparticle shapes.  A detailed description of the particle size determination method and calibrations are provided in Fig.~\ref{fig:fig_SiNP_size_determination}.

Figure~\ref{fig:fig2}c shows a darkfield microscopy image of a single Si nanoparticle stably trapped in the central region of the focused azimuthally polarized beam; the dashed circle outlines the trapping beam (its optical image was blocked with a notch filter as shown in Fig.~\ref{fig:fig_setup}). An \textit{in-situ} measurement of the single Si nanoparticle's scattering spectrum was performed (upper panel of Fig.~\ref{fig:fig2}d) to allow size determination of this and each single Si nanoparticle studied. The GLMT-calculated spectrum (lower panel of Fig.~\ref{fig:fig2}d) that matches the measured spectral features (peaks) establishes this particle's diameter to be 205~nm. This particle's MD resonance, measured and calculated in a water environment, is nearly maximally on-resonance with the 770~nm trapping laser wavelength, and the ratio of the optical magnetic dipole to electric dipole scattering is nearly maximized. 

The nanoscale localization\cite{qu2004nanometer} of single Si nanoparticles and the associated probability density functions (PDF) of different trapped Si nanoparticles along the radial direction (different colors) are shown in Figs.~\ref{fig:fig2}e and f, respectively. Particles of different sizes localize in different regions of the focused azimuthally polarized beam. Statistical analysis of experimental (upper) and simulation (lower) results demonstrate that the 205~nm dia. Si nanoparticle (blue points) is trapped at the beam center where the maximum longitudinally polarized magnetic field $H_z$ exists. This result is consistent with optical magnetic field trapping with a trapping force given by Eq.~\ref{eqn:dpfm}. By contrast, a 162~nm dia. Si nanoparticle (red points) is localized in the annular region of the focused azimuthally polarized beam, which is consistent with electric field trapping. This latter result is consistent with our expectation because the ED mode dominates the interaction with the electric field of the trapping laser via Eq~\ref{eqn:dpfe}, which is the well-established electric point-dipole model of electric light-matter interactions\cite{nieto2010optical}. 

Surprisingly, Figs.~\ref{fig:fig2}e and f also show that a large, 239~nm dia., Si nanoparticle (green) is strongly trapped at the center of the beam, with a narrower distribution in the radial position PDF! This result is unexpected since it starkly contrasts with that for the 162~nm dia. nanoparticle even though in both cases the interaction with their ED mode dominates the MD mode for 770~nm trapping laser light. Furthermore, as shown in Fig.~\ref{fig:fig_SCM_data_matrix_optical_properties}, the magnetic polarizability (\yl{and therefore induced magnetization}) is negative for the 239~nm dia. nanoparticle at 770~nm trapping laser wavelength. Therefore, the optical magnetic dipole gradient force of Eq.~\ref{eqn:dpfm} would be repulsive (see Fig.~\ref{fig:repulsive_magnetic_PD}), and the 239~nm dia. Si nanoparticle should be expelled from the center of the trap! Other Si nanoparticles of various sizes that we trapped and characterized showed results consistent with the three nanoparticles we discussed here in detail. See the Supplementary Information and Methods sections for more details about the experiments and simulations. These results clearly indicate that an additional, unaccounted-for optical force must exist to resolve the contradiction with expectations from a simple point-dipole model (of Eqs\ref{eqn:dpfe} and~\ref{eqn:dpfm}) and that this force dominates the electrodynamic forces acting on large Si nanoparticles. 

\begin{figure}[h!]
	\centering
	\hspace*{-0.5cm}
	\includegraphics[clip,trim=0cm 1cm 0cm 1cm,scale=.42]{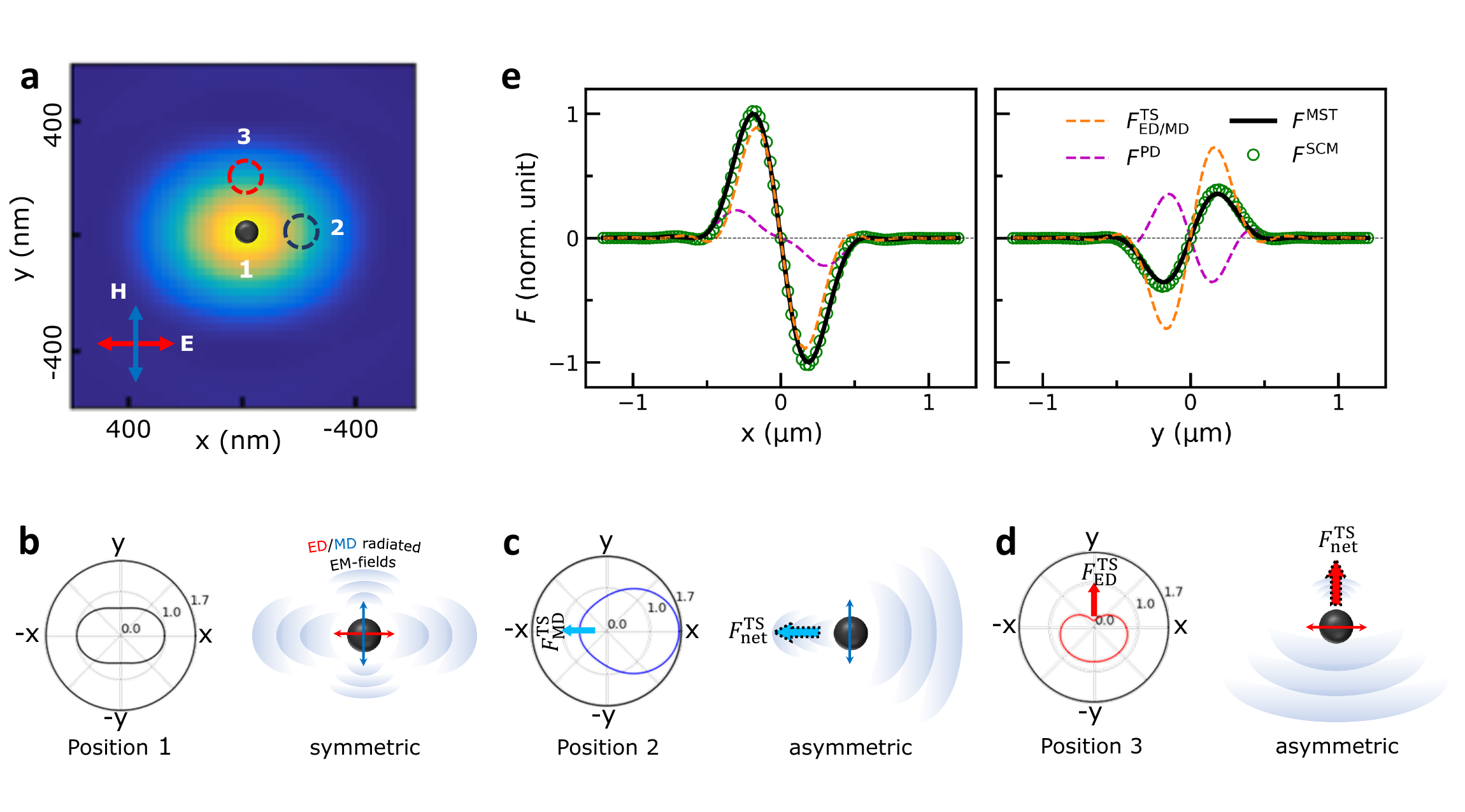}
	\caption{\textbf{Transverse scattering force(s) on Si nanoparticles due to the photonic Hall effect (PHE).} 
	\textbf{(a)} The PHE for the case of the dia.~205~nm dia. Si nanoparticle is demonstrated by characterizing its scattering properties at three designated positions (labeled 1, 2, and 3) in a linearly-polarized, tightly-focused Gaussian beam.	
        \yl{\textbf{(b)-(d)} Normalized far-field radiation patterns (left) of the Si nanoparticle at the three locations in \textbf{(a)} show that light scatters differently depending upon the polarization state of the optical beam and the local intensity gradients of the exciting electric and magnetic fields. These results define the force contributions from the PHE on the Si nanoparticle described in Eq.~\ref{eqn:tsf}. The schematic (right) describes the Mie transverse scattering (TS) from a Si nanoparticle optically excited near its electric or magnetic dipolar resonances. A particle at the center of the optical trapping beam manifests symmetric scattering that becomes asymmetric when the particle is away from the center . A net force $F^{\text{TS}}_{\text{net}}$ arises to balance the momentum carried by the directional scattering.  
        \textbf{(e)} Optical forces exerted on the Si nanoparticle calculated at various positions along the x- and y-directions in the Gaussian beam shown in \textbf{(a)} calculated by four methods: transverse scattering (TS); point-dipole (PD) approximation; Maxwell stress tensor (MST); and our scattering corrected model (SCM). Note, all forces are normalized to the maximum of the MST-calculated force.}
			}
	\label{fig:fig3}
\end{figure}

\subsection*{Transverse scattering (TS) force and photonic Hall effect}
Figure~\ref{fig:fig3} qualitatively and quantitatively demonstrates the directional forces associated with the transverse directional scattering of light. Figure~\ref{fig:fig3}b shows the results of GLMT simulations of a Si nanoparticle interacting with a focused Gaussian beam where the electric and magnetic fields are polarized along the in x- and y-directions, respectively. Placing the 205~nm dia. Si nanoparticle at three representative positions in the trapping beam, as shown in Fig.~\ref{fig:fig3}a (labelled 1, 2, and 3), results in the three far-field scattering patterns shown in Fig.~\ref{fig:fig3}b-d. The scattering pattern for the particle at the origin (position 1) displays mirror (and inversion) symmetry in the transverse plane so no net TS force is generated, as shown schematically in Fig.~\ref{fig:fig3}b. Note that for this size nanoparticle, the light scattering at 770~nm originates from both the ED and MD modes. \yl{The excited dipole moments align with their respective exciting fields and radiate electromagnetic fields in the transverse plane (see Fig.~\ref{fig:fig3}b).} Therefore, the 205~nm dia. Si nanoparticle scatters more intensely along the x-axis because of the large scattering cross-section associated with the optical magnetic dipole (see Fig.~\ref{fig:fig2}d). \yl{It is important to note that this additional optical force also exists for small particles (such as the 162~nm dia. Si nanoparticle), but its trivial contribution to the total force on the Si nanoparticle allows the point-dipole model to accurately describe the electrodynamic forces in this size regime (see Fig.~\ref{fig:fig_SCM_data_matrix}).}

The mirror symmetry of the scattering is lost when the particle is displaced from the central axis of the trapping beam. The x- and y-axes are the (reflection) symmetry axes for the electric and magnetic components of the Gaussian beam, respectively. When the particle is placed at position 2, its deviation from the (y-oriented) symmetry axis for the magnetic field leads to symmetry breaking of the MD-mediated scattering along the x-direction, as shown Fig.~\ref{fig:fig3}c. Similarly, the asymmetric ED-mediated scattering at position 3 occurs because of the particle's displacement from the (x-oriented) symmetry axis for the electric field, as shown Fig.~\ref{fig:fig3}c. However, these two configurations show different biases in scattering, with the ED mode scattering light along the electric field intensity gradient and the MD mode scattering light in the direction opposite to the magnetic field intensity gradient. A Si nanoparticle off-center in an electromagnetic beam with a (typical) Gaussian transverse intensity profile, where its electric and magnetic fields are not spatially segregated, scatters the incident photons into the opposite direction (inward or outward) depending upon the photons' linear polarization state \yl{(i.e., along the x- or y-directions)}. This contrarian behavior is attributed to the interference between the MD and the ED modes, as reflected in Eq.~\ref{eqn:tsf}. \yl{This polarization-dependent directional scattering is the photonic Hall effect and is essential for understanding the trapping behavior of Si nanoparticles.}

\yl{Following this reasoning, we propose that a \textit{transverse scattering} force must be added to the gradient force model described in Eqs.~\ref{eqn:dpfe},~\ref{eqn:dpfm}, and~\ref{eqn:PDA_tot} (in Methods).} 
Figure~\ref{fig:fig3} shows that the TS force results from directionally biased scattering; the particle's ``reaction'' balances the momentum carried by the scattered radiation, satisfying conservation of momentum. Therefore, how the nanoparticles scatter light, especially for near-~or on-resonance optical trapping, is an important and perhaps dominant force acting on the nanoparticles. As we show below, this scattering force on the Si nanoparticle arises from the PHE\cite{onoda2004hall, hosten2008observation, haefner2009spin}. Polarization-dependent directional light scattering from nanoparticles is already known in the photonic spin Hall effect (PSHE), which relies on: (i) the light's circular polarization state; and (ii) gradients in the (local) refractive index of materials\cite{haefner2009spin,rodriguez2010optical,shi2019enhanced}. In contrast, in the present case, due to the existence of both MD and ED resonances and their interaction as indicated in Eq.~\ref{eqn:tsf}, the PHE acting on the Si nanoparticles depends on: (i) the incident light's linear polarization state; and (ii) the intensity gradients of the electric and magnetic fields \yl{(more details in Sect.~\ref{SI_subsect:PHE} in Supplementary Information)}. 

The TS forces are a consequence of the biased scattering just described, and have magnetic and electric dipolar contributions, as evidenced by an analytical approximation:

\begin{equation}
	\label{eqn:tsf}
	F^{\text{TS-ana}}=F^{\text{TS}}_{\text{MD}}+F^{\text{TS}}_{\text{ED}}~\yl{\cong}~-\frac{k^3\mu_{\text{m}}}{12\pi\epsilon_{\text{m}}}\text{Im}\left[\alpha_{\text{e}}\alpha_{\text{m}}^{*}\right]\left(\cos^2\theta\nabla|U_B|^2-\sin^2\theta\nabla|U_E|^2\right),
\end{equation}

\noindent where $k$ denotes the wavevector in the surrounding medium with permittivity $\epsilon_{\text{m}}$ and permeability $\mu_{\text{m}}$; $U_{B}$ and $U_{E}$ represent the energy densities of the magnetic and electric fields of light, respectively; and the parameter $\theta$ is the angle between the electric field vector and the direction corresponding to the gradient of the respective field's energy density \yl{(see the derivation in Eqs.~\ref{eqn:tsf-ana_origin}-\ref{eqn:tsf-ana} in Methods and the quantification in Fig.~\ref{SI_subsect:TSF_analytical} in the Supplementary Information)}. A general rule can be deduced from Eq.~\ref{eqn:tsf}: \textit{MD-mediated TS forces ($F^{\text{TS}}_{\text{MD}}$) attract particles to magnetic field maxima, while ED-mediated TS forces ($F^{\text{TS}}_{\text{ED}}$) repel particles away from electric field maxima}. 

We verified the TS forces and their properties indicated in Eq.~\ref{eqn:tsf} by numerically solving for the net momentum carried by the (unevenly) scattered radiation and precisely quantify the TS forces with Eqs.~\ref{eqn:TSF} and~\ref{eqn:AmpFuncs} in Methods. \yl{These results are benchmarked against rigorous numerical MST ($F^{\text{MST}}$) calculations. Figure~\ref{fig:fig3}e shows that the TS forces ($F_{\text{ED/MD}}^{\text{TS}}$; orange curves) were numerically calculated for the particle at various positions along the electric (left, x) and magnetic (right, y) field axes in the Gaussian beam shown Fig.~\ref{fig:fig3}a.} The behavior of the scattering force in Fig.~\ref{fig:fig3}e highlights the anisotropic nature of the dipole-based scattering forces---i.e., being attractive and repulsive along the orthogonal axes.

The optical forces computed using the Maxwell stress tensor (MST, in black)\cite{novotny2012principles} or point-dipole model (PD, in green dash)\cite{novotny2012principles}, are also presented in addition to the numerically calculated TS forces. Comparing these forces demonstrates the significance of the TS forces when optical trapping occurs in the optical resonant regime (see Fig.~\ref{fig:fig_SCM_data_matrix}). The MST-computed forces are the most accurate determination of the actual optical forces experienced by the Si nanoparticle because the MST rigorously represents the particle's electromagnetic response by fully solving Maxwell's equations. The substantial discrepancies shown in Fig.~\ref{fig:fig3}e between the MST-computed forces and the PD-computed forces highlight the inadequacy of the conventional point-dipole model in describing the forces acting on the particle. In fact, the TS forces ($F^{\text{TS}}$) account for the forces omitted in the PD model. The resulting ``Scattering Corrected Model'' (SCM; green circles) \(F^{\text{SCM}}_{\text{total}}=F^{\text{PD}}+F^{\text{TS}}\) is suitable for understanding the electrodynamics of resonant optical trapping and optical magnetic contributions (see Methods and Supplementary Information).

\begin{figure}[H]
	\centering
	\hspace*{-0.5cm}
	\includegraphics[clip,trim=0cm 5cm 0cm 2cm,scale=.5]{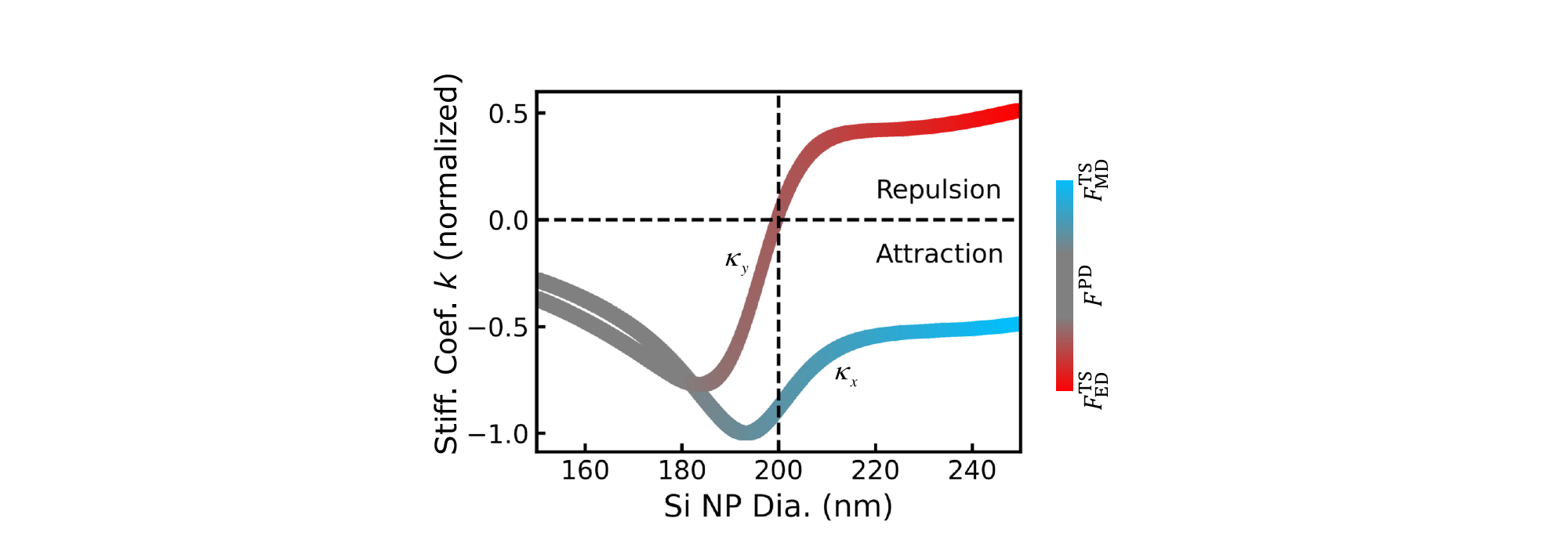}
	\caption{\textbf{Computed optical trap stiffness characterization.} 
		Trapping stiffness coefficients of the Si nanoparticle in the x- and y-directions in the Gaussian beam shown in Fig.~\ref{fig:fig3}a  as a function of nanoparticle diameter. The color is encoded to indicate the primary contribution transitioning from dipole force to transverse scattering force as the particle’s size increases. Positive values of $\kappa_y$ mean repulsion that pushes the particle away from the central area of the Gaussian beam in the y-direction.
	}
	\label{fig:fig4}
\end{figure}

\subsection*{Scattering corrected model of optical trapping forces}
The SCM explains the trapping behavior of Si nanoparticles in various optical beams including the focused azimuthally polarized beam employed here. Figure~\ref{fig:fig4} shows the trapping stiffness coefficients along the electric and magnetic field axes (as defined in Fig.~\ref{fig:fig3}a) as a function of particle dimensions. The encoded colors indicate the dominant contribution to the force on the Si nanoparticle and how these contributions change with Si nanoparticle diameter, from the dipole force ($F^{\text{PD}}$: gray) to the ED/MD-mediated TS forces ($F^{\text{TS}}_{\text{ED}}$/$F^{\text{TS}}_{\text{MD}}$: red/blue) in red/blue. The changes occur because the resonances shift to longer wavelengths with Si nanoparticle size thereby tuning on-resonance with the ED and MD modes (as seen in Fig.~\ref{fig:fig1}c). Two key observations arise from the results shown in Fig.~\ref{fig:fig4}. First, $|\kappa_x|$ reaches its minimum value while $|\kappa_y|$ is nearly zero near the optical magnetic dipole resonance (vertical black dashed line), indicating that particles in this size range are primarily influenced by the MD-mediated TS force. Second, $|\kappa_x|$ and $|\kappa_y|$ become comparable in magnitude but with opposite sign for larger particles. This means that large Si nanoparticles will simultaneously experience attraction and repulsion from the optical magnetic dipole and electric dipole-mediated TS forces in orthogonal directions.

\begin{figure}[H]
	\centering
	\hspace*{-0.6cm}
	\includegraphics[clip,trim=4cm 0cm 4cm 1cm,scale=.55]{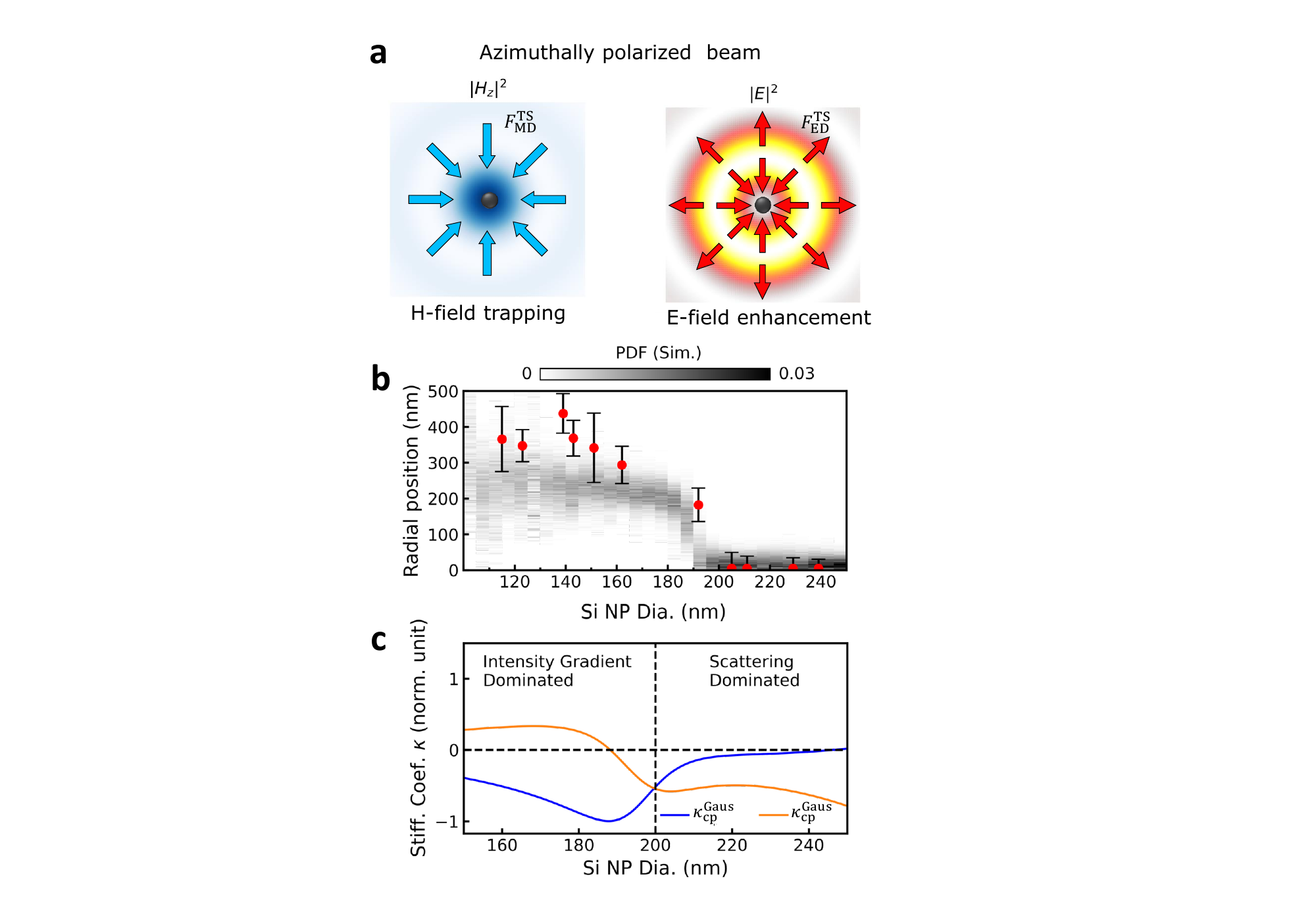}
	\caption{\textbf{Synopsis of experimental and simulated results of Si nanoparticle trapping in an azimuthally polarized beam.}
		\textbf{(a)} The mechanism for an azimuthally polarized beam stabilizing Si nanoparticles at its center principally relies on the magnetic field scattering force $F^{\text{TS}}_{\text{MD}}$ actuated by the $H_z$ component of the magnetic field. This inward-directed optical force drives Si nanoparticles to the most intense $H_z$ region (left panel). The electric scattering force $F^{\text{TS}}_{\text{ED}}$ actuated by the electric field $E$ enhances the trapping stability by repulsion (right panel). Note, this ``pushing force'' is bi-directional on each side of the ring profile; therefore, the pushing force itself will not be able to confine Si nanoparticles as firmly as the magnetic field scattering force $F^{\text{TS}}_{\text{MD}}$.
		\textbf{(b)} Radial distribution of Si nanoparticles trapped in the azimuthally polarized beam plotted as a function of their diameters, ranging from 100~nm to 250~nm in steps of 5~nm. Experimental radial distributions (red points with standard deviation error bars) of Si nanoparticles with selected dimensions are superimposed for comparison.
		\textbf{(c)} Comparison of the trap stiffness coefficient of different sized Si nanoparticles trapped in azimuthally $\kappa^{\text{AZM}}$ and circularly polarized Gaussian $\kappa^{\text{Gaus}}_{\text{CP}}$ beams provide guidance of beam selection for trapping particles in the two distinct particle size trapping regimes.
	}
	\label{fig:fig5}
\end{figure}

These two properties of the TS forces reveal the mechanism of optical magnetic field trapping. Figure~\ref{fig:fig5}a schematically decomposes the optical forces exerted on a Si nanoparticle under illumination from an azimuthally polarized beam. When the particle is excited near its optical magnetic dipole resonance, it is attracted toward the magnetic field maximum due to $F^{\text{TS}}_{\text{MD}}$ (left panel). \yl{The optical MD mode of larger Si nanoparticles is detuned to longer wavelengths relative to the (770~nm) trapping laser.} Therefore, the Si nanoparticles ($>$200~nm dia.) experience not only attractive forces due to optical magnetic fields through the magnetic TS force, but also a repulsion from $F^{\text{TS}}_{\text{ED}}$ from the radial intensity gradient of $E(r,\Phi)$ in azimuthally polarized optical beams. The latter pushes the Si nanoparticle towards the beam center (Fig.~\ref{fig:fig5}a right panel). This is why both the 205 nm and 239 nm Si nanoparticles (and others shown in Fig.~\ref{fig:fig5}b) were trapped at the location of the longitudinally polarized magnetic field, $H_{z}$. The 239~nm dia. Si nanoparticle shows greater confinement because $F^{\text{TS}}_{\text{ED}}$, the transverse scattering force, becomes larger for larger Si nanoparticles. \yl{More quantitative details of the transverse scattering force effects as well as intensity gradient forces are shown in Figs.~\ref{SI_subsect:TSF_analytical}-\ref{fig:force_analysis_239sinp}.}  

Figure~\ref{fig:fig5}b shows the radial position distribution (PDF) calculated as a function of particle diameter demonstrating how the optical trapping of Si nanoparticles depends on particle size. These results were obtained using GLMT-Langevin dynamics simulations (see Methods). The simulation (gray) and experimental (red points) results show an abrupt transition from electric field trapping to optical magnetic field trapping when the trapping laser is near/on-resonance with the magnetic dipole mode (Dia.=200 nm). \yl{Moreover, the decrease in the standard deviations of the distributions with increasing particle size shows that the trapping efficacy is enhanced. This result further confirms the validity of our force analysis and the key role of the photonic Hall effect.} 

\section*{Discussion and Conclusions}
\yl{We have demonstrated experimentally that optical magnetic field trapping of Si nanoparticles using azimuthally polarized beams. However, the mechanism is more complex than simply involving magnetic gradient forces as represented in Eq.~\ref{eqn:dpfm}. The failure to accurately explain the experimentally observed electrodynamic forces on the Si nanoparticles with a simple extension of the standard electric point-dipole model (Eq.~\ref{eqn:dpfe}) to consider magnetic dipole gradient forces (Eq.~\ref{eqn:dpfm} and Eq.~\ref{eqn:PDA_m} in Methods), led us to identify the ED- and MD-mediated transverse scattering forces and quantify their unique properties. The latter are a manifestation of the photonic Hall effect for Si nanoparticles. This novel realization required the development of the scattering (force) corrected model (Eq.~\ref{eqn:tsf} and Eq.~\ref{eqn:tsf-ana} in Methods), that determines the TS forces quantitatively (via numerical computation). The TS forces ($F^{\text{TS}}$) provide a missing element to the tradition point-dipole picture and enabled us to uncover optical magnetic aspects of optical trapping. This development gives a generalized description of electromagnetic trapping.}

\yl{The generalized description of optical trapping, including optical magnetic forces that this paper introduces, suggests opportunities that utilize the optical resonances of the trapping sample (material constituents). First, our results extend to other high index of refraction materials as shown in supplementary Figs.~\ref{fig:varMaterials}-\ref{fig:varMaterials_trapping}, for both visible and mid-infrared wavelengths. Second, our results contravene the notion that scattering forces are antithetical to stable optical trapping as suggested by Ashkin through the optical Earnshaw's theorem\cite{Ashkin1983stability}, especially given their divergence-free nature. As indicated in Eq.~\ref{eqn:tsf}, the TS forces' reliance on the gradients of the field energy density is fundamentally akin to the well-known dipole forces that rely on the gradient of light's intensity and are not divergenceless. This similarity becomes evident when examining the TS force fields generated in azimuthally polarized beams, as schematically demonstrated in Fig.~\ref{fig:fig5}a where a well-defined potential is created solely from scattering forces. Our electrodynamics-Langevin dynamics simulation results shown in Fig.~\ref{fig:fig5}c, which compare the trapping stiffnesses of differently sized Si nanoparticles trapped in an azimuthally polarized beam ($\kappa^{\text{Azm}}$) and a circularly polarized Gaussian beam ($\kappa_{\text{cp}}^{\text{Gaus}}$), reveal that while the Gaussian beam provides stable trapping of Si nanoparticles with small sizes (whose optical magnetic dipole is off-resonance with the trapping beam), its trapping force degrades dramatically as the particle becomes larger. In contrast, the optical magnetic field-mediated scattering is maximized in an azimuthally polarized beam when the trapping laser is maximally on-resonance with the MD mode.} This new insight paves the way for exploring mesoscopic quantum optomechanical systems that are currently limited by the use of red-detuned dipole traps, restricting their particle selection to those with small dimensions and low-index materials\cite{gieseler2021optical}.

The present work also creates a new opportunity to study many body effects and optical matter formation resulting from magnetic (vs.~electric) light-matter interactions. Additionally, we showed that under illumination of a nonuniform (intensity) linearly polarized beam, the dipole-governed scattering forces exhibit anisotropic transverse trapping forces that can be leveraged to design optical-driven mechanical systems for a variety of applications such as particle sorting and optical conveyors as well as optofluidic wells\cite{nan2018sorting}.

Finally, although we have extended the mathematical description of optical trapping and have demonstrated optical magnetic trapping with the scattering forces, there remains an open question: Can particles be trapped in the longitudinally polarized magnetic field $H_{z}$ solely by magnetic gradient forces? Answering this question is important since it will serve as direct evidence for the possibility of constructing red-detuned optical magnetic traps where only magnetic dipole forces are considered without scattering forces involved. \yl{This demonstration requires the trapped particles' electric and magnetic dipole resonances to be spectrally well-separated so that the strength of MD associated forces are large relative to ED forces and even larger than that of the Si nanoparticle we reported in this paper. We believe that this proof-of-principle experiment could be accomplished by utilizing meta-atoms\cite{manna2017selective} and/or high index materials with MD resonances in the mid-infrared region (see supplementary Fig.~\ref{fig:varMaterials}).}

\bibliography{scibib}
\bibliographystyle{naturemag}

\newpage

\input{Methods}

\section*{Acknowledgements}
We acknowledge support from the Vannevar Bush Faculty Fellowship program sponsored by the Basic Research Office of the Assistant Secretary of Defense for Research and Engineering, and a Suzuki Postdoctoral Fellowship Award to YL. We acknowledge support from the National Science Foundation (NSF) Grant No. ECCS-1809410, and we thank support from Illinois State University for support. We also acknowledge the University of Chicago Research Computing Center for providing the computational resources needed for this work.

\section*{Author contributions}
N.F.S. and Y.L. conceived the idea and experiments. Y.L. and E.V. designed the optical system, carrying out all the measurements, and analyzed the collected data along with S.N., under the supervision of N.F.S. and S.A.R. Y.L. established the scattering-corrected model to explain the scattering forces and identify their connection to the photonic Hall effect. J.P. developed our in-house software, MiePy, by which Y.L. and E.V. performed electrodynamics Langevin-dynamics simulations to interpret the experimental observations. M.P., U.M., and M.B. fabricated the Si nanoparticle materials. Y.L. and N.F.S wrote the manuscript with contributions from all co-authors. 

\section*{Competing interests}
The authors declare no competing interests.

\newpage
\input{SI}

\end{document}

%% file: Methods.tex
\section*{Methods}
\subsection*{Electrodynamics simulations of scattered EM fields and forces on Si nanoparticles}
\subsubsection*{\uline{Lorentz-Mie theory}}
The scattering cross-section spectrum of Si nanoparticles, as shown in Fig.~1c in the main text, was calculated using Lorentz-Mie theory (LMT) as expressed by\citesupp{hulst1981light}:

\begin{equation}
\label{eqn:LMT}
	C_{sca}=\frac{2\pi}{k^2}\sum_{n=1}^{\infty}\left(2n+1\right)[|a_n|^2+|b_n|^2]
\end{equation}

\noindent where $k$ is the wave vector in the surrounding medium. The complex-valued $a_n$ and $b_n$ denote the electric- and magnetic-associated scattering coefficients, respectively, with the subscript $n$ representing the order of multipolar modes. We use the same definitions throughout this Methods section and Supplementary Information unless stated otherwise.

\subsubsection*{\uline{Generalized Lorenz-Mie theory}}
Generalized Lorentz-Mie theory (GLMT) is an extended version of LMT that takes shapes of electromagnetic beams into account. This allows computing the field scattered by a particle located at an arbitrary position in an electromagnetic field\citesupp{gerard2011generalized}. We previously developed an open-source software package, MiePy\citesupp{parker2021miepy}, based on this method that enables computing the scattering fields of various-sized Si nanoparticles in both azimuthally polarized and (linearly or circularly polarized) Gaussian beams. In the \yl{GLMT framework}, an incident beam (Eq.~\ref{eqn:GLMT_inc}) and its resultant scattered fields (Eq.~\ref{eqn:GLMT_scat}) are, respectively, represented by electric- and magnetic-associated vector spherical harmonic wave (VSHW) functions $N_{mn}^{(1),(3)}$ and $M_{mn}^{(1),(3)}$\citesupp{parker2020collective}:  

\begin{equation}
\label{eqn:GLMT_inc}
	\textib{E}_{\text{inc}}=-\sum_{n=1}^{L_{\text{max}}}\sum_{m=-n}^{n}iE_{mn}[p_{mn}N_{mn}^{(1)}+q_{mn}M_{mn}^{(1)}]
\end{equation}

\begin{equation}
\label{eqn:GLMT_scat}
\textib{E}_{\text{sca}}=\sum_{n=1}^{L_{\text{max}}}\sum_{m=-n}^{n}iE_{mn}[a_np_{mn}N_{mn}^{(3)}+b_nq_{mn}M_{mn}^{(3)}]
\end{equation}

\noindent where $n$ and $m$, respectively, denote multipolar and azimuthal orders; $L_{\text{max}}$ indicates the maximal multipolar order to which the field equations need to be expanded; $E_{mn}$ are normalization constants, $p_{mn}$ and $q_{mn}$ are the beam shape coefficients determined by the incident field's property as defined in the following subsection. Both the $a_n$ and $b_n$ are the (regular) scattering coefficients determined by the LMT given in Eq.~\eqref{eqn:LMT}. The superscript index, $\left(1\right)$ and $\left(3\right)$, of the VSHW terms commonly refer to the incident and scattered fields, respectively\citesupp{parker2020collective}.

\subsubsection*{\uline{Tightly focused beam modeling}}
To precisely model the tightly focused laser beam in the simulations, the excitation fields studied (i.e., azimuthal and Gaussian beams) were represented in the far-field angular spectrum representation of Hermite-Gaussian modes in spherical coordinates\citesupp{parker2020collective},

\begin{equation}
\label{eqn:HG}
\begin{split}
	\textib{E}_{l,m}^{\text{HG}}\left(\theta,\phi\right)&=(-i)^{l+m}H_l\left(\frac{kw_0}{\sqrt{2}}\tan\theta\cos\phi\right)H_m\left(\frac{kw_0}{\sqrt{2}}\tan\theta\sin\phi\right) \\
	&\quad \exp[\left(-kw_0\tan\theta/2\right)^2],
\end{split}
\end{equation}

\noindent where $w_0$ is the beam waist, and $H_{l,m}$ denotes Hermite polynomials of order $l$ and $m$. Therefore, the azimuthally polarized beam is described as a linear combination of the first order (0,1) and (1,0) Hermite-Gaussian modes 

\begin{equation}
\label{eqn:azm}
\textib{E}_{\infty}^{\text{Azm}}=\hat{x}\textib{E}_{0,1}^{\text{HG}}-\hat{y}\textib{E}_{1,0}^{\text{HG}},
\end{equation} 

\noindent while Gaussian beams are the lowest order (0,0) Hermite-Gaussian mode

\begin{equation}
\label{eqn:gau}
	\textib{E}_{\infty}^{\text{Gaus}}=\hat{p}\textib{E}_{0,0}^{\text{HG}},
\end{equation}

\noindent where $\hat{p}$ indicates the polarization direction.

With the incident field in the form of the far-field angular spectrum representation, the determination of the beam shape coefficients, $p_{mn}$ and $q_{mn}$, in Eq.~\ref{eqn:GLMT_inc} can be accomplished using an integral projection method\citesupp{parker2020collective}. The respective focal fields are computed using a Fourier transform, which is conveniently expressed in cylindrical coordinates as Ref\citesupp{parker2020collective}   

\begin{equation}
\label{eqn:AngSpecRep}
	\textib{E}\left(\rho,\varphi,z\right)=\frac{ikf\mathrm{e}^{-ikf}}{2\pi}\int\displaylimits_0^{\theta_{\text{max}}}\int\displaylimits_0^{2\pi}\textib{E}_{\infty}^{\text{Azm,Gaus}}\mathrm{e}^{ikz\cos\theta}\mathrm{e}^{ik\rho\sin\theta\cos\left(\phi-\varphi\right)}\sin\theta \, \mathrm{d}\phi \, \mathrm{d}\theta,
\end{equation}

\noindent where the term before the integral, which is associated with the focal length $f$ and medium-mediated wave vector $k$, serves as the focal field's amplitude. The maximum latitude angle $\theta_{\text{max}}$ limits the angular range of the focused beam, and equivalently defines the NA ($=n\sin\theta_{\text{max}}$, where $n$ is the medium's refractive index) of the microscope objective producing the focal field. The NA-dependent intensity ratios of field components of the focused azimuthally polarized beam shown in Fig.~1b were obtained by varying $\theta_{\text{max}}$ with corresponding value of the NA from 0.4 to 1.29. 

\subsubsection*{\uline{Maxwell stress tensor}}
The total optical force that determines the Si nanoparticles' motion in the electromagnetic fields is evaluated by integrating a Maxwell stress tensor (MST), $\overset{\text{\tiny$\leftrightarrow$}}{\textib{T}}$, over a closed surface, $\Omega$, enclosing the investigated particle\citesupp{novotny2012principles}

\begin{equation}
\label{eqn:F_MST}
	\langle \textib{F}^{\text{MST}}\rangle=\int_{\Omega}\langle \overset{\text{\tiny$\leftrightarrow$}}{\textib{T}}\rangle\cdot\textib{n}~\mathrm{d}a,
\end{equation}

\begin{equation}
\label{eqn:MST}
	\langle \overset{\text{\tiny$\leftrightarrow$}}{\textib{T}}\rangle=\frac{1}{2}\text{Re}\left[\epsilon_{\text{m}}\textib{E}\otimes\textib{E}^*+\mu_{\text{m}}\textib{H}\otimes\textib{H}^*-\frac{1}{2}\left(\epsilon_{\text{m}}E^2+\mu_{\text{m}}H^2\right)\overset{\text{\tiny$\leftrightarrow$}}{\textit{\textbf{I}}}\right],
\end{equation}

\noindent where $\langle\ldots\rangle$ indicates the time average; $\textbf{n}$ is the unit vector perpendicular to an infinitesimal surface element $\mathrm{d}a$; $\otimes$ is the vector outer product; the superscript asterisk indicates the conjugate fields; $\overset{\text{\tiny$\leftrightarrow$}}{\textit{\textbf{I}}}$ is the unit tensor; and $\epsilon_{\text{m}}$ and $\mu_{\text{m}}$ denote the permittivity and permeability of the surrounding medium, respectively. It is important to understand that the fields ($\textib{E}$ and $\textib{H}$) are the superposition of the incident ($\textib{E}_{\text{inc}}$ and $\textib{H}_{\text{inc}}$) and the scattered fields ($\textib{E}_{\text{sca}}$ and $\textib{H}_{\text{sca}}$) for self-consistency. The full consideration of incoming and outgoing fields explicitly renders the momentum transfer from the photons to the particles\citesupp{novotny2012principles}. \yl{Therefore, the MST-calculated force represents the complete electromagnetic-field-induced force acting on the particles and is the accurate ground truth that the forces in the point-dipole (PD) approximation and scattering-corrected model (SCM) are compared to. Details are provided in the next sections.} 

\subsubsection*{\uline{Point-dipole (PD) approximation}}
In the PD model, the investigated Si nanoparticles are assumed to be Rayleigh particles that simultaneously exhibit electric and magnetic dipole resonances that allow interaction with the incident trapping laser's electromagnetic field and, consequently, manifest two field-induced dipole moments. Both the electric and magnetic components of the illuminating light thereby have contributions to the overall mechanical force exerted by the electromagnetic field on the Si nanoparticles. Therefore, it is appropriate to cast the mechanical force as

\begin{equation}
\label{eqn:PDA_tot}
	\langle\textib{F}^{\text{PD}}\rangle=\langle\textib{F}_{\text{e}}^{\text{PD}}\rangle+\langle\textib{F}_{\text{m}}^{\text{PD}}\rangle.
\end{equation}

\noindent The analytical expressions for each individual field-determined force can be expressed as\citesupp{novotny2012principles}:

\begin{equation}
\label{eqn:PDA_e}
\langle\textib{F}_{\text{e}}^{\text{PD}}\rangle=\frac{\alpha'_{\text{e}}}{4}\nabla|\textib{E}_{\text{inc}}|^2+\frac{\alpha''_{\text{e}}}{2}|\textib{E}_{\text{inc}}|^2\nabla\Phi,
\end{equation}
\noindent and
\begin{equation}
\label{eqn:PDA_m}
\langle\textib{F}_{\text{m}}^{\text{PD}}\rangle=\frac{\alpha'_{\text{m}}}{4}\nabla|\textib{H}_{\text{inc}}|^2+\frac{\alpha''_{\text{m}}}{2}|\textib{H}_{\text{inc}}|^2\nabla\Phi,
\end{equation}

\noindent where the subscripts $\text{e}$ and $\text{m}$ refer to the electric- and magnetic-field relevant parameters, respectively, and $\alpha_{j}=\alpha'_{j}+i\alpha''_{j}$ denotes the particle's complex electric ($j=\text{e}$) and magnetic ($j=\text{m}$) polarizabilities. \yl{Note that an alternative form of the PD model, particularly suitable for trapping light with arbitrary polarizations, can be found elsewhere\citesupp{ruffner2013comment}.} The dipole forces only consider incident fields as indicated in Eqs.~\ref{eqn:PDA_e} and~\ref{eqn:PDA_m}. The first terms in Eqs.~\ref{eqn:PDA_e} and~\ref{eqn:PDA_m} are the intensity gradient forces (IGF) and the second terms are the radiation pressure that involves the phase gradient, $\nabla\Phi$. As indicated by Eqs.~\ref{eqn:PDA_e} and \ref{eqn:PDA_m}, only the IGF influences the particle's motion at the focal plane (where $\nabla\Phi$ = 0) in the PD model, while the radiation pressure is a longitudinal force driving particle along the incident beam's propagation direction. 

Since the investigated Si nanoparticles have dimensions beyond (larger than) the Rayleigh regime, the PD theory needs to be extended by defining the particle's polarizabilities in terms of the Mie scattering coefficients\citesupp{moroz2010non,selmke2014energy,belov2003condition},

\begin{equation}
\label{eqn:pol_e}
\alpha_{\text{e}}=i\frac{6\pi\epsilon_{\text{NP}}}{k^3}a_1,
\end{equation}
\noindent and
\begin{equation}
\label{eqn:pol_m}
\alpha_{\text{m}}=i\frac{6\pi}{\mu_{\text{NP}}k^3}b_1,
\end{equation}

\noindent where $\epsilon_{\text{NP}}$ and $\mu_{\text{NP}}$ are the permittivity and permeability of the particle, respectively. The first order Mie scattering coefficients, $a_1$ and $b_1$, are associated with the electric and magnetic dipole resonances, respectively (see Eq.~\ref{eqn:LMT}). 

However, the Mie-extended PD treatment is not sufficient to describe forces on Si nanoparticles when excited at their ED and MD resonances. Furthermore, the interaction between the MD \yl{\underline{and}} ED becomes non-negligible and results in momentum transfer from the incident to the resultant scattering fields that is not captured within the point-dipole approximation as expressed in Eqs.~\ref{eqn:PDA_e} and \ref{eqn:PDA_m}.

\subsubsection*{\uline{Scattering-corrected model (SCM)}}
\yl{We developed the SCM to capture the transverse scattering force acting on Si nanoparticles, particularly when their dipolar scattering strength is very significant, i.e., when on-resonance with the MD and ED modes. The SCM-calculated force, $\textib{F}^{\text{SCM}}$, is constituted of two forces: the dipole force and the transverse scattering (TS) force}

\begin{equation}
\label{eqn:SCM}
	\langle\textib{F}^{\text{SCM}}\rangle=\langle\textib{F}^{\text{PD}}\rangle+\langle\textib{F}^{\text{TS}}\rangle.
\end{equation} 

\noindent In the present study, the electromagnetic-field-driven motion of the Si nanoparticles is restricted to the focal plane, and therefore only the transverse component of the scattering force is taken into account. The dipole force in Eq.~\ref{eqn:SCM} determines the intensity gradient force; i.e., the first terms in Eq.~\ref{eqn:PDA_e} and \ref{eqn:PDA_m}.

The TS force is non-conservative and created by the transverse component of the field scattered by a Si nanoparticle. In the scattering process, a portion of the incident beam's energy is transferred to the Si nanoparticle, exciting its dipole mode. This energy is subsequently dissipated through absorption and scattering. Momentum conservation requires that light scattered by the Si nanoparticle should result in a recoil (reaction) of the particle in an opposite direction, and the recoil force should be proportional to the intensity of the scattered field. 

\yl{We derive an expression to quantify the net TS force. Our derivation builds on theoretical studies on optical scattering forces\citesupp{gordon1973radiation,rohrbach2001optical,rohrbach2002trapping,rohrbach2004reply}.} The TS force in a specific direction $\bm{\hat{r}}$ with orientation ($0^{\circ}$, $\varphi$) is obtained by integrating the intensity of the scattered field, $I_{scat}(\theta,\phi)$, over polar $\theta$ and azimuthal $\phi$ angles:

\begin{align}
    \label{eqn:TSF}
	\textib{F}^{\text{TS}}(\varphi)=\bm{\hat{r}}\int\displaylimits_{0}^{2\pi}\int\displaylimits_{0}^{\pi}\frac{2\pi\epsilon_{0}}{k^2}\textit{I}_{scat}(\theta,\phi)\cos(\phi-\varphi)\sin\theta \, \mathrm{d}\theta \, \mathrm{d}\phi,
\end{align}

\noindent where $I_{scat}(\theta,\phi)=|S_1(\theta,\phi)|^2+|S_2(\theta,\phi)|^2$. The amplitude functions $\left[ S_1(\theta,\phi), S_2(\theta,\phi)\right]$ define the amplitudes of the scattered fields in any directions. Based on GLMT theory and notation, they are expressed as: 

\begin{align}
\label{eqn:AmpFuncs}
	S_1(\theta,\phi) &= \sum_{n=1}^{L_{\text{max}}}\sum_{m=-n}^{n}E_{mn}\left[ a_np_{mn}\pi_{mn}(\cos\theta)+b_nq_{mn}\tau_{mn}(\cos\theta)\right]e^{im\phi} \\
	S_2(\theta,\phi) &= \sum_{n=1}^{L_{\text{max}}}\sum_{m=-n}^{n}E_{mn}\left[ a_np_{mn}\tau_{mn}(\cos\theta)+b_nq_{mn}\pi_{mn}(\cos\theta)\right]e^{im\phi}, \nonumber
\end{align}

\noindent where $\pi_{mn}(\cos\theta)$ and $\tau_{mn}(\cos\theta)$ are generalized Legendre functions\citesupp{hulst1981light}. The rest of the parameters ($ E_{mn}$, $a_n$, $b_n$, $p_{mn}$, $q_{mn}$) have been previously introduced in Eqs~\ref{eqn:LMT}-\ref{eqn:GLMT_scat}.

\yl{As shown in Figs. 3b-d of the main text, when a Si nanoparticle is placed at the center of a focused linearly polarized beam (e.g., Gaussian beam), its dipole radiation maintains mirror symmetry; the exciting field does not exert a net TS force on the particle. However, when mirror symmetry is broken by displacing the nanoparticle from the  center of the beam, a net TS force correspondingly arises on the particle that can be determined quantitatively using Eq.~\ref{eqn:TSF}.}

\subsubsection*{\uline{Analytical approximation of the transverse scattering force}}
\yl{Numerical computation of Eq~\ref{eqn:TSF} allows for a precise quantification of TS forces. Yet, a deeper physical insight into the TS force arising from the magnetic field $H_z$ and would benefit from an analytical expression. Moreover, the photonic Hall effect for Si nanoparticles that we have introduced can be better understood. Given the dual-dipole resonance (i.e., ED and MD) of Si nanoparticles, their scattering behavior should be governed by an effective polarizability accounting for both the electric ($\alpha_{\text{e}}$) and magnetic ($\alpha_{\text{m}}$) polarizabilities. This process is analogous to two-body systems where two optically bound nanoparticles can alter the effective polarizability of dipole-dipole structures\citesupp{sule2017rotation} through the interplay between the induced electric dipole responses. This dipolar interactions between the induced electric and magnetic modes has been theoretically proposed by M. Nieto-Vesperinas \textit{et al.} in regard to optical forces acting on small magnetodielectric particles\citesupp{nieto2010optical}.} 

Based on the physical parameter defined here, we reproduce the essential part of the expression derived by M. Nieto-Vesperinas \textit{et al.}\citesupp{nieto2010optical} of the TS force:

\begin{equation}
\begin{split}
\label{eqn:tsf-ana_origin}
	F^{\text{TS-ana}} & = -\frac{k^3\mu_{\text{m}}}{6\pi\epsilon_{\text{m}}}\text{Im}\left[\alpha_{\text{e}}\alpha_{\text{m}}^{*}\right]\text{Im}\left[\textib{E}\times \textib{B}^{*}\right] \\
 & = -\frac{k^3\mu_{\text{m}}}{12\pi\epsilon_{\text{m}}}\text{Im}\left[\alpha_{\text{e}}\alpha_{\text{m}}^{*}\right]\left[\textib{E}^{*}\times\left(\nabla\times\textib{E}\right)+\textib{B}^{*}\times\left(\nabla\times\textib{B}\right)\right].
\end{split}
\end{equation}
\noindent In the first line, we intentionally omit a term associated with $\text{Re}\left[\textib{E}\times \textib{B}^{*}\right]$ in the original expression, Eq.~44 of Ref\citesupp{nieto2010optical}, since it represents the longitudinal force along the laser beam propagation direction. Obtaining the second line from the first line of Eq.~\ref{eqn:tsf-ana_origin} requires the identity: 

\begin{equation}
\label{eqn:tsf-ana_equality}
    \text{Im}\left[\textib{E}\times \textib{B}^{*}\right]=\frac{1}{2}\left[\textib{E}^{*}\times\left(\nabla\times\textib{E}\right)+c^2\textib{B}^{*}\times\left(\nabla\times\textib{B}\right)\right].
\end{equation}

Using the paraxial approximation and cylindrical coordinate system convention [with unit direction vectors ($\vu{\text{e}}_{\rho}$, $\vu{\text{e}}_{\theta}$, $\vu{\text{e}}_{z}$)], the electric and magnetic fields of a linearly polarized Gaussian beam are:
\begin{equation}
\begin{split}
\label{eqn:para_EB_fields}
    \textib{E} & = u_{E}\cos{\theta}\vu{\text{e}}_{\rho}-u_{E}\sin\theta\vu{\text{e}}_{\theta}; \\
    \textib{B} & = -\frac{u_{B}}{c}\sin{\theta}\vu{\text{e}}_{\rho}-\frac{u_{B}}{c}\cos\theta\vu{\text{e}}_{\theta}.
\end{split}
\end{equation}
\noindent where $u_{E/B}$ is the fundamental transverse Gaussian mode (including phasor $e^{\left[i\left(kz-\omega t\right)\right]}$) describing the electric and magnetic field amplitudes. The parameter  $\theta$ denotes the electric field polarization direction that ranges from $0^{\circ}$ (x-polarization) to $90^{\circ}$ (y-polarization). 

Substituting Eq.~\ref{eqn:para_EB_fields} into the second line of  Eq.~\ref{eqn:tsf-ana_origin}, yields the TS force:  

\begin{equation}
\label{eqn:tsf-ana}
	F^{\text{TS-ana}}=-\frac{k^3\mu_{\text{m}}}{12\pi\epsilon_{\text{m}}}\text{Im}\left[\alpha_{\text{e}}\alpha_{\text{m}}^{*}\right]\left(\cos^2\theta\nabla|U_B|^2-\sin^2\theta\nabla|U_E|^2\right),
\end{equation}

\noindent where $|U_E|$ and $|U_B|$ denote the energy densities of the electric and magnetic fields of light, respectively. \yl{This new development of Eq.~\ref{eqn:tsf-ana} points out that the (mechanism of) the TS force depends on both the electric field polarization and the magnetic field polarization directions.}


For example, assume a general trapping condition where a Si nanoparticle is exposed to a paraxial Gaussian beam with x-polarization (electric field). Equation.~\ref{eqn:tsf-ana} implies that when the particle's position is aligned along the polarization direction of the electric field ($\theta=0^{\circ}$), the TS force is mainly associated with the gradient of the magnetic field, $|U_B|$. Moreover, the TS force then acts perpendicular to the magnetic field polarization direction. Conversely, when the particle's position is aligned along the polarization direction of the magnetic field ($\theta=90^{\circ}$), the TS force is solely determined by the gradient of the electric field, $|U_E|$, and acts perpendicular to the electric field polarization direction. Moreover, the opposite signs before the trigonometric functions best exemplify the anisotropy of dipole-mediated TS forces. Therefore, the electric and magnetic field-actuated TS forces are always orthogonal to each other and exhibit opposite behaviors, where if one is attractive, then the other must be repulsive. The final determination of attraction or repulsion requires considering the sign of the crossing term, $\text{Im}\left[\alpha_{\text{e}}\alpha_{\text{m}}^{*}\right]$.

Equation.~\ref{eqn:tsf-ana} can be simplified by unifying the energy densities $\left(U_E=U_B=U\right)$ since the electric and magnetic fields of light share the same spatial profile in the paraxial approximation. Hence, Eq.~\ref{eqn:tsf-ana} can be rewritten:

\begin{equation}
	\label{eqn:tsf-ana-para}
	F^{\text{TS-ana}}=-\frac{k^3\mu_{\text{m}}}{12\pi\epsilon_{\text{m}}}\text{Im}\left[\alpha_{\text{e}}\alpha_{\text{m}}^{*}\right]\cos2\theta\nabla|U|^2.
\end{equation} 

\noindent Equation~\ref{eqn:tsf-ana-para} shares the exact same mathematical structure as one reported by X. Xu \textit{et al.}\citesupp{xu2020kerker}. Once again, in comparison, Eq.~\ref{eqn:tsf-ana} reveals more physical insights that allow differentiating the contributions of both the electric field and the magnetic field to the TS forces.

Although this analytical method is derived from assuming a paraxial Gaussian beam, it can still provide a meaningful approximation for tightly focused beams of light as well as structured electromagnetic fields, thus complementing the rigorous numerical approach represented by Eq.~\ref{eqn:TSF} with a physical interpretation. Its applicability in tightly focused Gaussian and azimuthal beam is investigated in Sect.~\ref{SI_subsect:TSF_analytical}-\ref{SI_subsect:anaF_239nmSiNP}.

\subsubsection*{\uline{Langevin dynamic simulations}}
To have a better understanding of Si nanoparticles' dynamics observed in our experiments, GMMT electrodynamics-Langevin dynamics (ED-LD) simulations were undertaken using software, termed MiePy, developed in the Scherer group\citesupp{parker2021miepy}. MiePy was designed for simulations of optical matter systems and nanoparticles in optical traps so we could quantitatively capture the experimental conditions\citesupp{parker2020collective}. Our ED-LD simulations are parameterized to closely agree with the temporal dynamics of particles in the azimuthally polarized electromagnetic field in a fluid (water) environment at various temperatures. The driven motion of a single Si particle in water medium can be described by the overdamped Langevin equation\citesupp{parker2020optical}

\begin{equation}
\label{eqn:LD}
	\frac{\mathrm{d}\textib{r}}{\mathrm{d}t}=\frac{1}{\gamma}\textib{F}+\sqrt{\frac{2k_BT}{\gamma}}\text{\boldmath$\eta$},
\end{equation}

\noindent where \textib{r} denotes the position of the particle at time $t$, $\gamma=6\pi\mu R$ is the friction coefficient of the (spherical) particle with radius $R$ in fluid of viscosity $\mu=8\times 10^{-4}~\text{Pa}\cdot\text{s}$. The electrodynamic force \textib{F} is numerically determined by the MST using Eq.~\ref{eqn:F_MST}. The second term, including the Boltzmann constant $k_B$ and the medium temperature $T=300~\text{K}$ as well as a delta-correlated random normal vector \text{\boldmath$\eta$} with mean 0 and variance 1, accounts for the Brownian motion of the particle in room temperature fluid. 

\yl{The parameters that were adjusted to obtain the results shown in Fig.~\ref{fig:fig2} in the main text, once the Si nanoparticle diameter was chosen, is the power of the optical beam and tightness of focus of the simulated optical beam at the focal plane.}

\subsubsection*{\uline{Force and potential energy maps}}
The potential energy maps shown in Fig.~\ref{fig:fig1}d in the main text were acquired by uniformly pixelating a trapping plane with grid size of 10~nm and evaluating the optical forces at the defined spatial positions. The optical force acting on the Si nanoparticles of interest was calculated using the MST in Eq.~\ref{eqn:F_MST}. Integrating the force along a predefined path $s$ can determine a relative potential energy at the path's terminal location $s_{\text{t}}$ by: 

\begin{equation}
\label{eqn:PEM}
	V(s_{\text{t}})=-\int\displaylimits_{s_{\text{i}}}^{s_{\text{t}}}\textib{F}(s) \, \mathrm{d}s.
\end{equation}

\noindent There is no restriction to the selection of the path as well as its initial position $s_{\text{i}}$. Yet, once the initial position is defined, it should be treated as a reference value for the energy and work calculations. A potential energy map is then obtained after collecting relative potential energies at all coordinates. For the potential energy maps shown in Fig.~\ref{fig:fig1}d, their lower left corners were routinely chosen as the reference point with potential energy of 0.

\bibliographystylesupp{naturemag}
\bibliographysupp{scibib}

%% file: SI.tex
\topmargin 0.0cm
\oddsidemargin 0.2cm
\textwidth 16cm 
\textheight 21cm
\footskip 1.0cm

\renewcommand{\thesubsection}{S\arabic{subsection}}

\setcounter{figure}{0}
\renewcommand{\thefigure}{S\arabic{figure}}

\setcounter{equation}{0}
\renewcommand{\theequation}{S\arabic{equation}}

\title{\centering{\huge Supplementary Information for}\\
\vspace{1cm}
\centering{\huge Optical trapping with optical magnetic field and photonic Hall effect forces}\\
\vspace{1cm}}

\tableofcontents

\pagebreak


\subsection{Synthesis of Si nanoparticles}
\label{SI_subsect:synthesis}

\begin{figure}[H]
	\centering
	\includegraphics[clip,trim=1.8cm 2.5cm 2cm 0.5cm,scale=0.55]{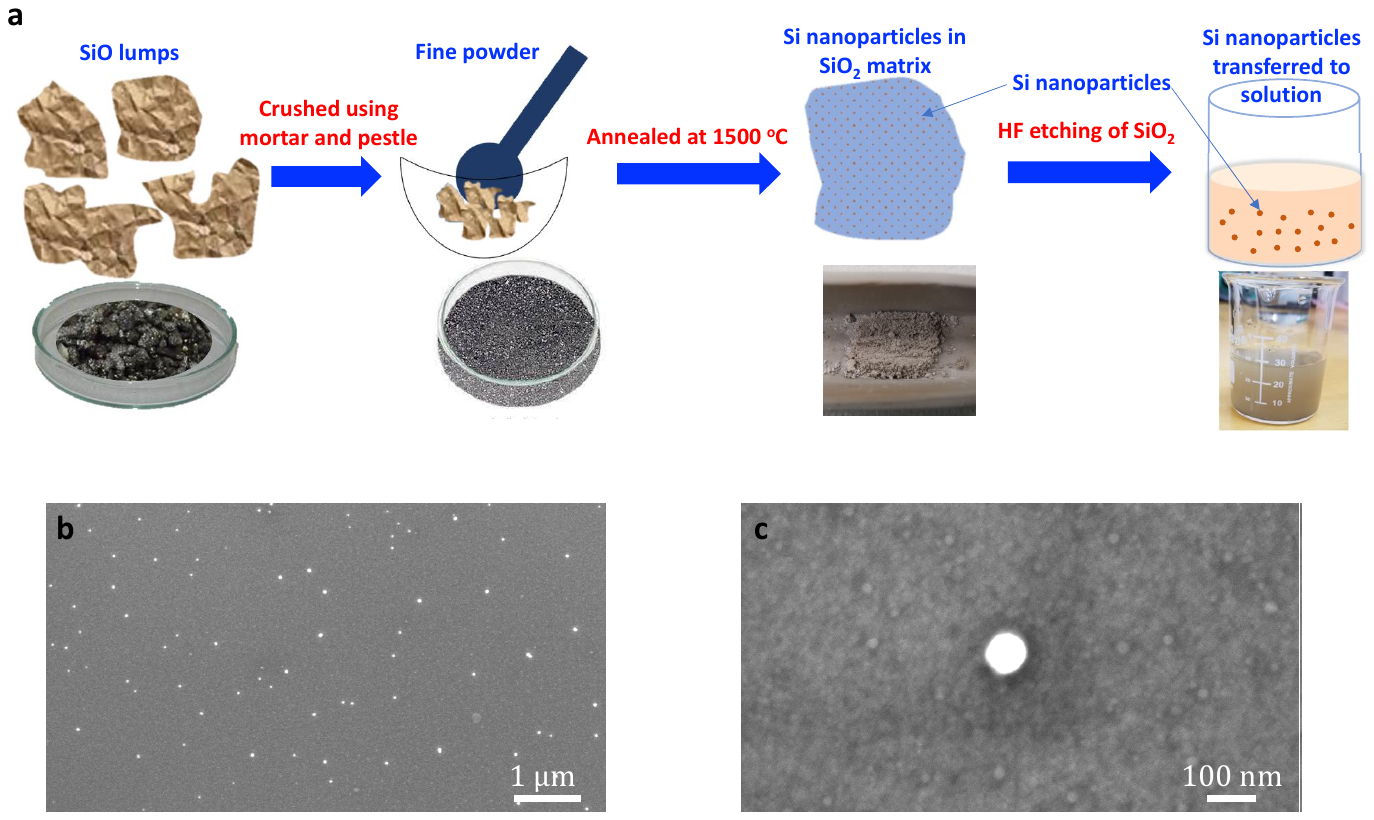}
	\caption{\textbf{Preparation of Si nanoparticle samples.}
		\textbf{(a)} The schematic outlines the procedure of Si nanoparticle synthesis.
		\textbf{(b)} SEM micrograph displays the fabricated Si nanoparticles deposited on a conductive substrate (ITO on glass). 
  \textbf{(c)} A magnified view of a single Si nanoparticle particle (160~nm dia.). 
	}
	\label{fig:SI1}
\end{figure}

The procedure used for Si nanoparticle synthesis shown in Fig.~\ref{fig:SI1}a follows a previously reported method\citesuppl{sugimoto2020mie}. Lumps of 99.99\% purity 3-10 mm in size SiO (Sigma Aldrich) were crushed to powder and annealed at 1500 °C in an inert gas atmosphere for 30 minutes using an MTI GSL-1700X tube furnace to promote the solid-state disproportionation of SiO into Si and SiO$_2$. This was followed by etching the SiO$_2$ matrices in a concentrated hydrofluoric acid (HF) solution (48\% from Sigma Aldrich) for one hour, which resulted in the extraction of free Si nanoparticles. The extracted particles were transferred to water after several rinses through centrifugal filtration (AccuSpin 8C centrifuge by Fisher Scientific), followed by 2-minutes of ultrasonication using a Fisher Scientific FB505 sonic dismembrator. 

\newpage
\subsection{Size characterization of Si nanoparticle samples}
\label{SI_subsect:size_characterization}

\begin{figure}[H]
	\centering
        \hspace*{-1cm}
	\includegraphics[clip,trim=3cm 5.5cm 5cm 3cm,scale=.6]{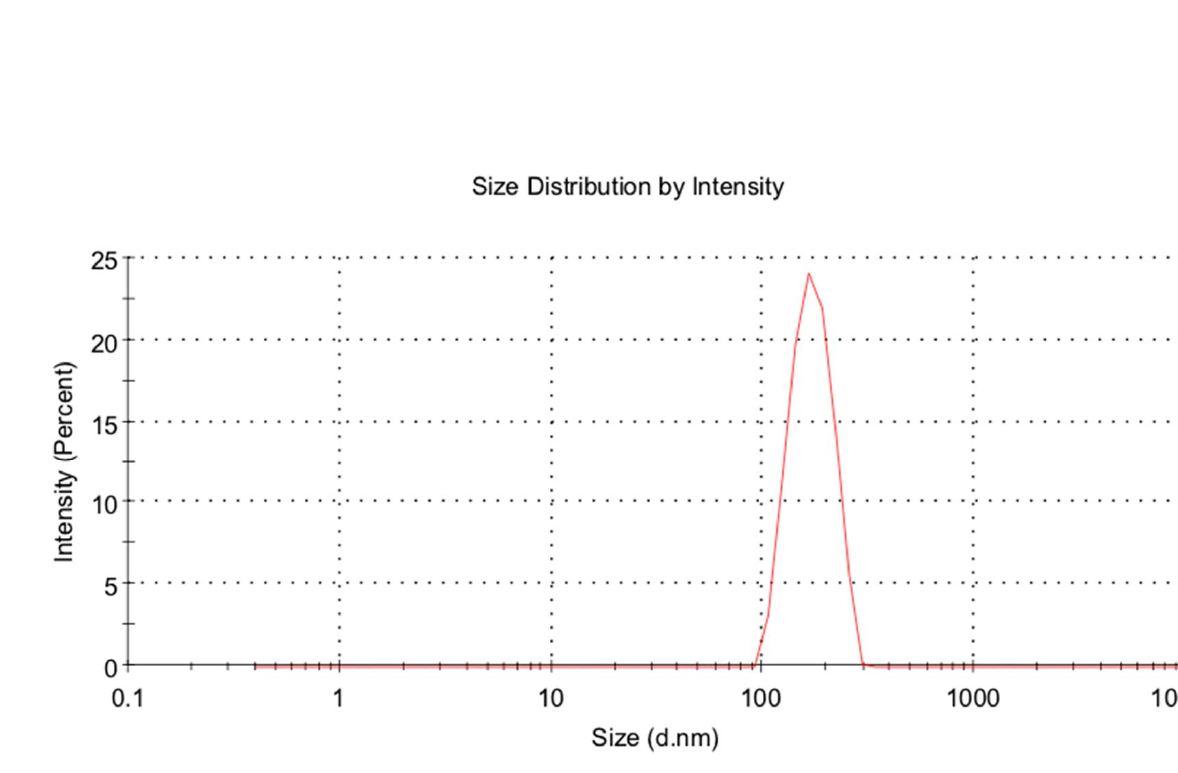}
	\caption{\textbf{Size distribution of Si nanoparticles from dynamics light scattering (DLS) measurements.}
		The synthesized Si nanoparticle sample exhibited a broad size range, spanning from 100~nm to 300~nm dia. (hydrodynamics), with a peak at approximately 200~nm diameter. Note the logarithmic size scale.
	}
	\label{fig:SI2}
\end{figure}

Dynamic light scattering measurements (DLS) were performed to characterize the polydispersity of the as-synthesized Si nanoparticles. In the measurements, an aqueous colloidal suspension of Si nanoparticles in water was sonicated for 3 minutes; a 1 mL aliquot was extracted immediately after sonication and transferred to a cuvette that was loaded into a Malvern Zetasizer Nano ZS for DLS measurement. Multiple DLS measurements were conducted. A representative outcome shown in Fig.~\ref{fig:SI2} demonstrates the particle sizes are in the range of 100~nm to 300~nm dia. The size distribution exhibits a log-normal profile with its peak at 200~nm diameter. The log-normal distribution means that the preponderance of particles are smaller than 200~nm in diameter. 

In addition, zeta potential measurements were undertaken to assess the stability of the Si nanoparticles in the colloidal suspension. A 1~mL sample was created in the same manner as just described and measured using the Malvern Zetasizer Nano ZS. The resulting representative zeta potential was -32.0~mV.

Figures~\ref{fig:SI1}b and c as well as Fig.~\ref{fig:fig2}b of the main text demonstrate the high morphological quality of the synthesized Si nanoparticles as determined with scanning electron microscopy (SEM, Carl Zeiss Merlin) measurements. The sample was prepared for measurement by  drop-casting that created a sparse distribution of Si nanoparticles on an indium tin oxide (ITO) coated coverslip (SPI Supplies\textsuperscript{\textregistered}). For higher resolution measurements, the Si nanoparticles were dispersed on a glass coverslip by drop-casting and transferred into a high resolution sputter coater (Cressington 208HR) for the deposition of a uniform Pt/Pd layer. A thin metal layer (thickness 5$\pm$0.2~nm) was deposited on the Si nanoparticles (and glass substrate) by carefully controlling the vacuum pressure (0.08 mbar) of the coater.

\newpage

\subsection{Experimental setup}

\begin{figure}[H]
	\centering
	\hspace*{0cm}
	\includegraphics[clip,trim=4.2cm 4cm 4cm 4cm,scale=.9]{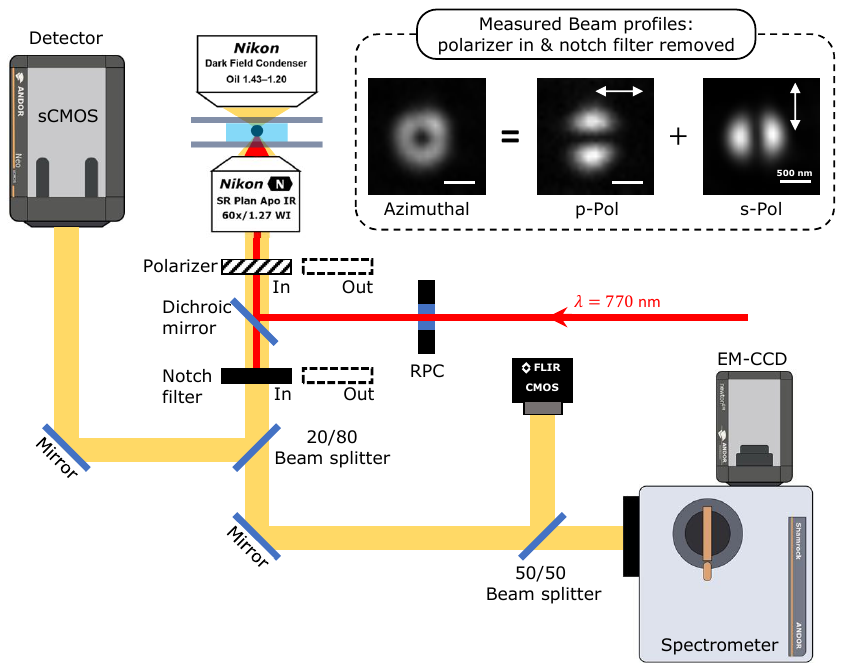}
	\caption{\textbf{Schematic of the optical setup for optical trapping of Si nanoparticles using an azimuthally polarized vector beam.}
		The main optical components used in the trapping experiment are shown. The relay (4f) lenses employed in the beam paths to the sCMOS detector and spectrometer are excluded for simplicity. The radial-polarization converter (RPC) was used to transform a collimated linearly polarized \yl{Gaussian} laser beam (red line) into the azimuthal polarization state. The inset shows the intensity distribution of the focused azimuthally polarized beam \yl{(left most)} at the trapping plane measured with the sCMOS detector. This azimuthally polarized beam's intensity profile was obtained by linearly combining the two intensity distributions shown in the inset, which represent the azimuthal beam passing through the polarizer with orthogonal polarization orientations (p- and s-pol), as indicated by the white arrows. Both the polarizer and notch filter can be moved in and out (as depicted by the dashed rectangle) of the optical path depending on the data acquisition requirements. 
	}
	\label{fig:fig_setup}
\end{figure}

\label{SI_subsect:experimental_setup}
\subsubsection*{\uline{Optical trap setup}}

Figure~\ref{fig:fig_setup} shows the optical setup used in the optical trapping experiments of Si nanoparticles using a focused azimuthally polarized laser beam. A collimated CW Ti-Sapphire \yl{laser beam} (at 770 nm wavelength) that was used as the trapping light source passed though a commercially available radial-polarization converter (ARCopix). This device converts the beam from a linear polarization to an azimuthal polarization state. The underlying principle of the polarization conversion is described elsewhere\citesuppl{stalder1996linearly}. The azimuthally polarized beam \yl{entered} the back port of an inverted optical microscope (Nikon, ECLIPSE Ti) and was focused using a 60X water-immersion microscope objective (Nikon CFI SR Plan Apo IR 60X) with numerical aperture (NA) of 1.27 into a water-filled sample chamber, constructed with a $\sim$120~$\mu$m thick SecureSeal\texttrademark imaging spacer (GRACE BIO\textendash LABS) sandwiched between a $\sim$127~$\mu$m thick microscope coverslip (bottom) and a $\sim$1~mm thick microscope slide (top, Fisher Scientific\texttrademark). Note that the glass slide is electrostatically charged with the same sign as the Si nanoparticles to create electrostatic repulsion to offset radiation pressure acting on them. 

This sample chamber was mounted on a closed-loop 3D (xyz) piezo electric transducer (PET) translation stage (Mad City Labs). The 3D PET stage was primarily used to  augment the microscope's motorized adjustment of the focal plane, and also to implement x, y translation to ascertain if Si  nanoparticles were stuck to the \yl{upper} glass surface of the sample chamber. Lateral translation also allowed calibration of the magnification of the microscope optics. 

The laser power was 170 mW before entering the rear pupil of the objective; some additional loss occurs due to the \textless100\% transmission through the objective.

\subsubsection*{\uline{Particle nanoscale localization measurements}}
Darkfield microscopy images of spectrally  broadband incoherent light from the high NA condenser of the inverted optical microscope scattered from the trapped Si nanoparticles were recorded with an sCMOS array detector (ANDOR, Neo). As shown in Fig.~\ref{fig:fig_setup}, 20\% of the incoherent scattered light was directed to the sCMOS array detector, while the remaining 80\% was transmitted through the beam splitter toward a spectrometer and a CMOS color (RGB) detector. A 50/50 beam splitter directs 40\% of the total scattered light to an imaging spectrometer and electron multiplying (EM) CCD array detector (ANDOR, Kymera 193i \& Newton\textsuperscript{EM}) that allows recording high-quality spectra. The remaining 40\% was imaged onto a CMOS detector (FLIR, Grasshopper 3) enabling simultaneous recording of RGB color images of single trapped Si nanoparticles. A representative image is shown in Fig.~\ref{fig:fig2}d in the main text.

The recorded single nanoparticle images (see Fig.~\ref{fig:fig2}c in the main text) with the sCMOS array detector (with an effective pixel size of 75~nm \yl{that is determined} from the physical pixel size of 6.5~\textmu m and 86X magnification that was determined by piezostage calibration) allows implementing nanoscale localization analysis of the pixellated image and fitting to a 2D Gaussian function\citesuppl{selmke2014energy,qu2004nanometer}. The localization precision depends inversely \yl{on} the square root of the number of detected photons (ignoring background). Si 150-250 dia. nanoparticles scatter light strongly so the localization precision is a few nanometers.
When investigating the trapping behavior of Si nanoparticles in the azimuthally polarized electromagnetic field, the polarizer was removed, and both the notch (Semrock Stopline\textsuperscript{\sffamily\textregistered}, 785~nm) and edge (Semrock BrightLine\textsuperscript{\sffamily\textregistered} 694~nm/SP) filters were inserted in the optical path to block the trapping laser light from being measured by the sCMOS detector. The incoherent scattered light darkfield microscopy images of the Si nanoparticles were recorded by the sCMOS detector at 675~fps with 1~ms exposure time. The imaging field of view was typically set to be 30 by 30 pixels (2.25~\textmu m~$\times$ 2.25~\textmu m). 
In addition, the color images of the trapped Si nanoparticles simultaneously obtained by the CMOS sensor (FLIR) allow real-time estimation of the nanoparticles' size by judging the principal color of their scattering light.

\subsubsection*{\uline{Azimuthal beam alignment and imaging}}
In the process of azimuthal beam alignment, both the notch and edge filters were removed from the optical path so that the trapping laser can be monitored by the sCMOS detector. Due to the azimuthal polarization of the trapping beam, the alignment was achieved by positioning the polarizer at appropriate angles and observing the dumbbell-like beam profiles at the two orthogonal polarization states, i.e., p- and s-polarization, as shown in the inset of Fig.~\ref{fig:fig_setup}. The annular-profile of the full azimuthally polarized beam was obtained by merging the two linearly polarized images of the beam shown in the inset of Fig.~\ref{fig:fig_setup}. The beam alignment was checked for each trapping experiment and its image recorded to allow identifying the azimuthal beam's location for data (localization) analysis. This allows creating overlays of the beam image and the nanoscale localizations of the Si nanoparticles as shown in Fig.~\ref{fig:fig2}e in the main text.

\subsubsection*{\uline{In-situ scattering measurements}}
Single Si nanoparticles were captured in the focused azimuthally polarized beam and localized to different regions of the focused beam, either at its center for large size nanoparticles or in its annular region for smaller nanoparticles. It was necessary to determine the size of each Si nanoparticle in order to \yl{establish} the trapping mechanism. Since the scattering spectra (as shown in Fig.~\ref{fig:fig2}d in the main text) are structured, they could be used for \textit{in-situ} size determination. As shown in Fig.~\ref{fig:fig_setup}, the \textit{in-situ} spectral measurements were carried out by focusing the incoherent light scattered from the optically trapped particle through a $\sim$0.35~mm wide slit onto the 150~l/mm grating of the spectrograph (ANDOR, Kymera 193i) and detecting the spectrum with an EMCCD (ANDOR, Newton\textsuperscript{EM}). Since the charged Si nanoparticles are repelled from the charged glass coverslip forming the top of the sample chamber, the scattering measurements were performed with a 25\% increase in the trapping laser power increasing the radiation pressure acting on the particle. This resulted in increased mechanical friction between the particle and the glass slide surface of the sample chamber. The increased friction (and stronger optical trapping forces) attenuated the Brownian motion of the particle such that the particle was firmly held in place and aligned with the spectrometer's slit during a 10-second-long spectral measurement process.  
\newpage

\subsection{\yl{Scattering and absorption cross-sections} of Si nanoparticles}
\label{SI_subsect:thermal_effect}

\begin{figure}[H]
	\centering
	\hspace*{0.4cm}
	\includegraphics[clip,trim=8.cm 4.5cm 0cm 4cm,scale=0.9]{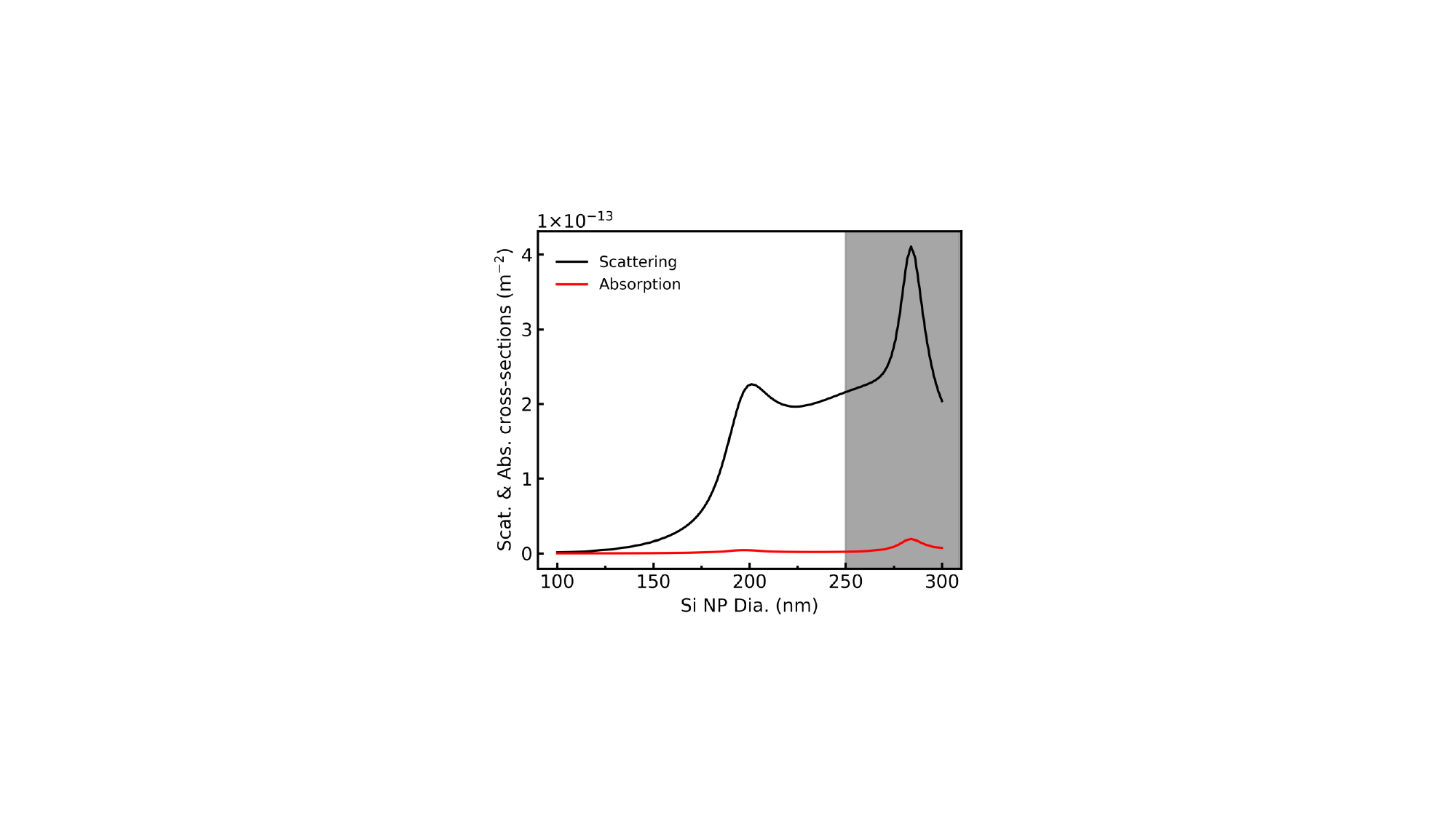}
	\caption{\textbf{Lorentz-Mie theory (LMT) calculated scattering and absorption spectra of a Si nanoparticle with various diameters at the trapping wavelength of 770~nm.} A very small feature is observed in the absorption of 770~nm light of the trapping laser that is associated with magnetic dipole resonance maximum at 200~nm dia. However, its magnitude is negligibly small in relation to the scattering. Note, the other feature near 280~nm dia. is from the magnetic quadrupole mode, which is in the gray shaded region that is out of the nanoparticle size range of interest as explained in Fig.~\ref{fig:fig1} in the main text.	
	}
	\label{fig:SiNP_apt_spectrum}
\end{figure}
Relative to the scattering cross-section, the calculated \yl{absorption cross-section} of 770~nm light by Si nanoparticles remains consistently close to zero throughout the nanoparticle size range of interest (within the white-colored \yl{region} of Fig.~\ref{fig:SiNP_apt_spectrum}), which reflects a minuscule absorbance of a Si nanoparticle illuminated at 770~nm. \yl{Even with the almost imperceptible increase in absorbance associated with} the magnetic dipole resonance for a 200~nm dia. Si nanoparticle, \yl{we conclude that} thermal \yl{heating of the aqueous solution will be} inconsequential \yl{based on our prior experience with metallic (Ag and Au) nanoparticles\citesuppl{figliozzi2017driven}.} \yl{Therefore, local temperature changes} are not considered in the present study.
\newpage

\subsection{Optical forces due to \yl{higher-order} resonances}
\label{SI_subsect:optical_force_high_order}

\begin{figure}[H]
	\centering
        \hspace*{2.8cm}
	\includegraphics[clip,trim=8.5cm 6.8cm 0cm 6.5cm,scale=1]{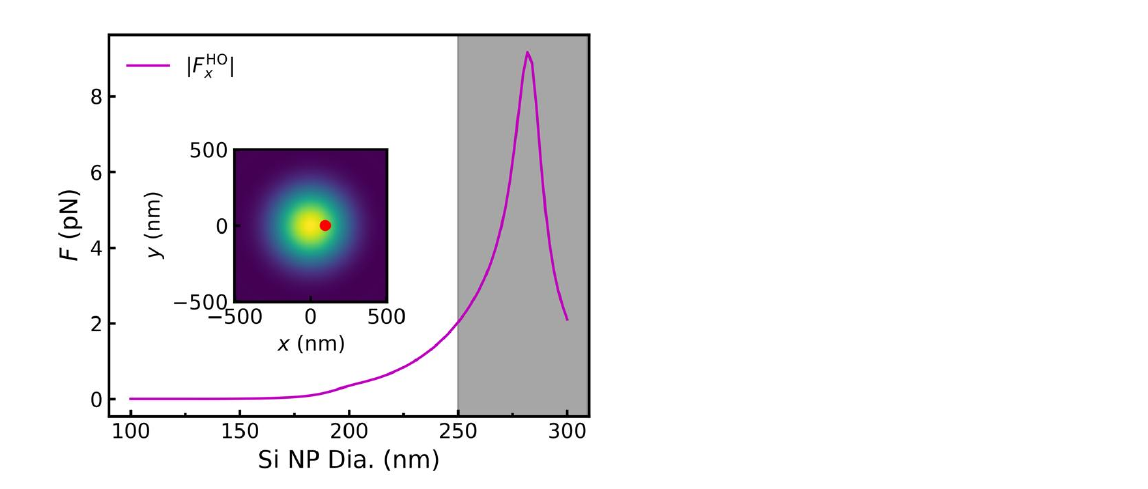}
	\caption{\textbf{Optical force arising from \yl{higher-order Mie resonances of} Si nanoparticles.}
		The force resulting exclusively from higher-order (beyond the first order \yl{dipolar}) optical resonances is evaluated for Si nanoparticles with dimensions ranging from 100~nm to 300~nm dia. Specifically, this is evaluated when the Si nanoparticle is positioned at a distance of 1/2 beam width away from the center of a circularly polarized Gaussian beam, as shown in the inset. The red dot indicates where the particle \yl{is situated} in the focused Gaussian beam.
	}
	\label{fig:high-order_force}
\end{figure}

As illustrated by the curve, the optical force \yl{in a Gaussian beam} due exclusively to \yl{higher-order} (mainly from the magnetic quadrupole) resonances gradually \yl{becomes important} with increasing size of the Si nanoparticle. This force becomes pronounced for sizes larger than 250~nm. In order to concentrate on dipole-governed electrodynamics, our study is limited to \yl{Si nanoparticles smaller than} 250~nm diameter.    
\newpage

\subsection{Calibration of scattering measurements for Si nanoparticles size determination}
\label{SI_subsect:SiNP_size_measurements}

\begin{figure}[H]
	\centering
	\hspace*{-0.6cm}
	\includegraphics[clip,trim=3cm 2cm 3cm 2cm,scale=.55]{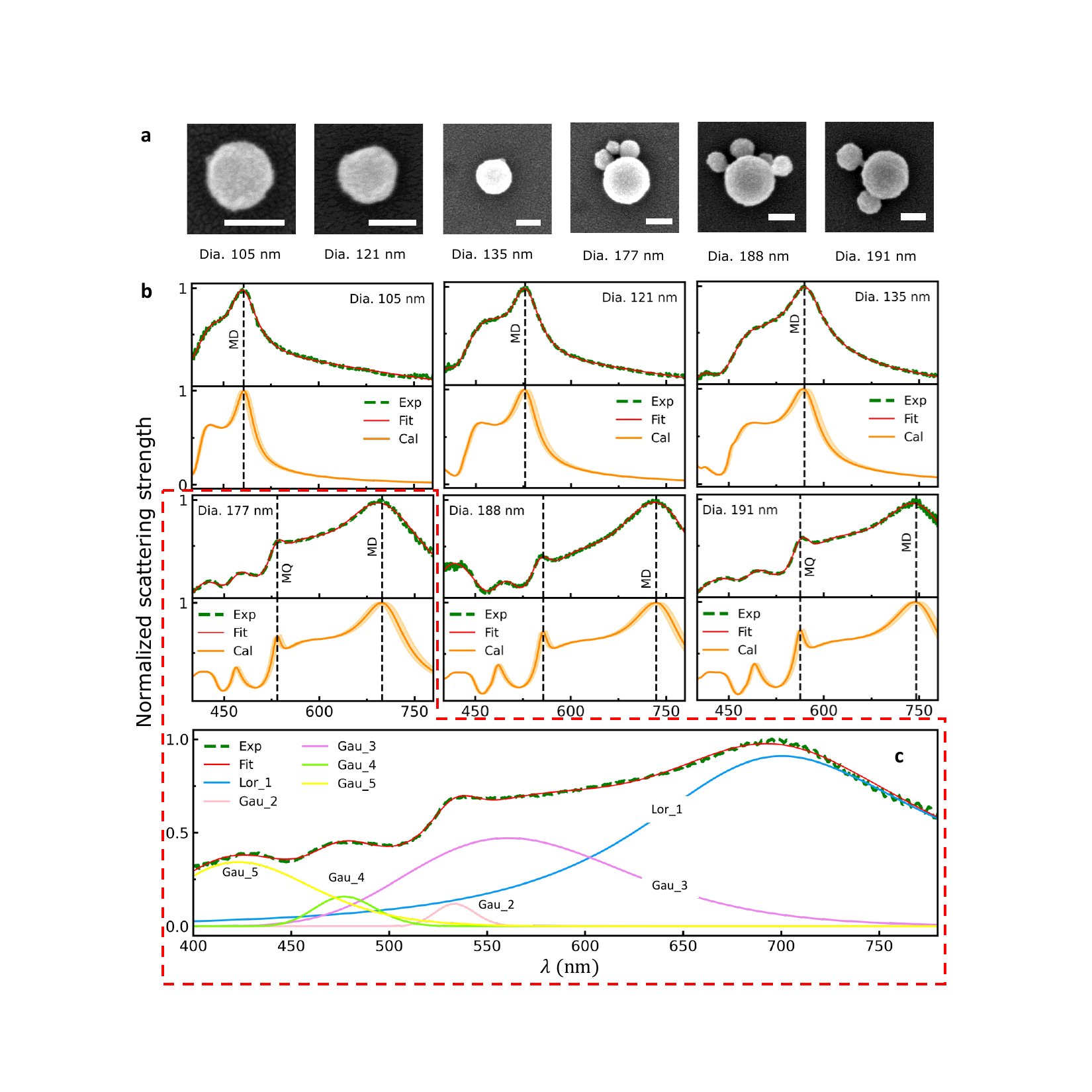}
	\caption{\textbf{Si \yl{nanoparticle} size determination by comparison of the measured and LMT-calculated scattering spectra.} \textbf{(a)} shows the SEM micrographs of various Si nanoparticles.
 \textbf{(b)} shows the spectra of Si nanoparticles with diameters ranging from small to large corresponding to the spectra in \textbf{(a)}. \yl{The} green dashed and red solid curves represent the experimental and corresponding fitted data, respectively. The orange curves with yellow shadow denote the LMT-calculated data with an error ($\pm$2~nm) according to the well-distinguished magnetic dipole (MD) and magnetic quadrupole (MQ) resonances marked by the black dashed lines. 
		Panel \textbf{(c)} is an example of the curve fitting specifically the experimental scattering spectrum of 177~nm dia. Si nanoparticle. The fitted curve is \yl{constituted of} five spectral features including one Lorentz (Lor) and four Gaussian (Gau) functions.
	}
	\label{fig:fig_SiNP_size_determination}
\end{figure}

The trapping behavior of a single Si nanoparticle under illumination of an azimuthally polarized beam \yl{depends on} the particle's size since it directly determines both the magnitude and relative dominance of the electric and magnetic dipole resonances excited. Understanding the \yl{size dependence} requires an accurate determination of each Si nanoparticle's size while being held by the electromagnetic field of the trapping laser. \textit{In-situ} scattering by the Si nanoparticles is most suitable and not disruptive of optical trapping. We calibrated the accuracy of this approach by comparing experimental scattering spectra and scanning electron microscopy (SEM) \yl{measurements} of single Si nanoparticles we intentionally deposited on the glass slide as shown in Fig.~\ref{fig:fig_SiNP_size_determination}a.



More than 40 Si nanoparticles with a large range of sizes were dispersed and stabilized on the surface of a microscope slide using a drop-casting method. The prepared sample was initially assessed with spectral scattering measurements, where it was incorporated into a water-filled chamber to mimic the conditions of the optical trapping environment. Therefore, the scattering spectra of the particles are chosen with typical dimensions as shown in Fig.~\ref{fig:fig_SiNP_size_determination}a that are representative of the ones obtained in the laser trapping experiments. 

The experimental scattering spectra (green dashed curves in Fig.~\ref{fig:fig_SiNP_size_determination}b) were compared with LMT-based calculations to establish the particles' sizes. The resonance features (peaks) observed in the experimental spectra (e.g., the magnetic dipole and quadrupole resonance peaks marked by the vertical black dashed lines) are well reproduced at the same spectral locations in the calculated spectra (orange solid curves in Fig.~\ref{fig:fig_SiNP_size_determination}b). In addition, a fitting model \yl{composed of a series of} Lorentz and Gaussian functions was developed to facilitate the identification of these experimentally acquired scattering signatures. The spectral resonance fitting functions, expressed in terms of angular frequency $\omega$, are:


\begin{equation}
	\label{eqn:LorOsc}
	Lor(\omega)=\frac{A\omega\gamma_b}{\omega_0^2-\omega^2-i\gamma_b\omega}
\end{equation}

\begin{equation}
	\label{eqn:GauOsc}
	Gau(\omega)=Ae^{-(\frac{\omega-\omega_0}{\gamma_b})^2}-Ae^{-(\frac{\omega+\omega_0}{\gamma_b})^2}
\end{equation}

\noindent where $A$, $\omega_0$, and $\gamma_b$ are the fitting parameters representing the amplitude, the center frequency, and the broadening of the resonances, respectively. 

\yl{Figure~\ref{fig:fig_SiNP_size_determination}c shows a detailed example of curve fitting that identifies five resonances for the Si nanoparticle with a diameter of 177~nm.} Here, $\omega_0$ serves as the spectral position of each peak. Specifically, \textit{Lor}\_1, \textit{Gau}\_2, and \textit{Gau}\_3 are mainly associated with the magnetic dipole, magnetic quadrupole, and electric dipole, respectively, while \textit{Gau}\_4 and \textit{Gau}\_5 are associated with higher-order resonances of the Si nanoparticle. It should be noted that the fitting functions used here are only for the purpose of determining the spectral positions of these resonance features, rather than ascertaining the true dielectric tensors of the Si material. 

The fitting process employs the Levenberg-Marguardt algorithm that minimizes the mean square error (MSE). The MSE is defined in terms of the deviations of the fitted $SC_i^{\text{fit}}$ and measured $SC_i^{\text{Exp}}$ data sets as given by:

\begin{equation}
	\label{eqn:MSE}
	MSE=\sqrt{\frac{1}{2N-M}\sum_{i=1}^{N}\left(\frac{SC_i^{\text{fit}}-SC_i^{\text{Exp}}}{\sigma_{SC,i}^{\text{Exp}}} \right)^2},
\end{equation}

\noindent where N is the sampling number (we use N=200 here), M is the total number of fitting model parameters, and $\sigma_{SC,i}^{\text{Exp}}$ are the standard deviations of the experimental data points. This fitting model, after analyzing all collected experimental data, reveals a nearly consistent standard deviation of $\pm$2~nm for the best-determined diameters. \yl{This standard deviation is represented in} the LMT-calculated spectra as yellow shadows in Fig.~\ref{fig:fig_SiNP_size_determination}a. 

After the particle sizes were characterized via the optical scattering measurement, the same particles were subsequently measured by SEM and visualized in Fig.~\ref{fig:fig_SiNP_size_determination}a. It is important to note that the small Si nanoparticles accumulated near the principle ones (for 177~nm, 188~nm, and 191~nm) when the sample was removed from the water-filled chamber and dried in a vacuum chamber. The surface tenson of the evaporating water could easily drag smaller particles, which usually have weaker adhesion to the surface, to the proximity of the larger ones. Therefore, the presence of these satellites in the SEM micrographs did not alter the \yl{scattering spectra and affect the sizes of the principle Si nanoparticles measured by SEM and the validation of particle size determination by comparing} the measured and LMT-calculated \yl{scattering spectra}. This agreement demonstrates the reliability and high precision of the size characterization with the \textit{in-situ} scattering measurements of the Si nanoparticles in an optical trap. 

\newpage

\subsection{Repulsive magnetic dipole force for large Si nanoparticles}
\label{SI_subsect:Repulsive_MDF}

\begin{figure}[H]
	\centering
        \hspace*{0.2cm}
	\includegraphics[clip,trim=1cm 2.5cm 0cm 2cm,scale=.52]{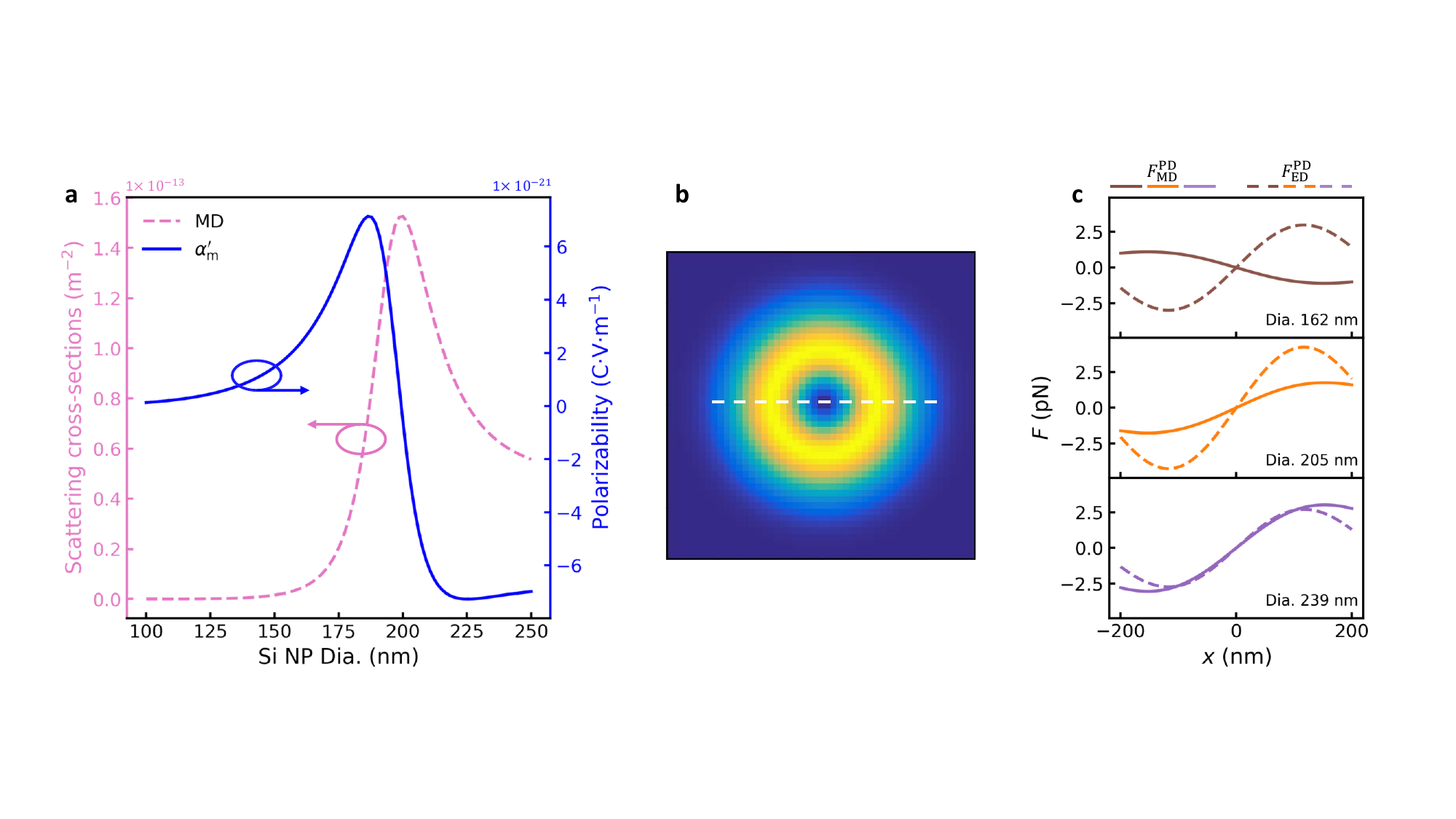}
	\caption{\textbf{Scattering cross-section, polarizability of and magnetic dipole force on Si nanoparticles trapped in an azimuthal beam.}
		\textbf{(a)} Magnetic dipole induced scattering cross-section (MD) and the real part of the magnetic polarizability ($\alpha'_{\text{m}}$) are computed as a function of Si nanoparticles size.
		\textbf{(b)} The azimuthal beam profile is rendered with a white dashed line crossing the beam center. That dashed line indicates the path along which the Si nanoparticles are scanned to evaluate the dipole forces they experience at each position.
		\textbf{(c)} Magnetic dipole forces $F_{\text{MD}}^{\text{PD}}$ (solid curves) are computed for three Si nanoparticles with dimensions corresponding to the experimental ones discussed in the main text. In addition, electric dipole forces $F_{\text{ED}}^{\text{PD}}$ (dashed curves) are also included for comparison.
	}
	\label{fig:repulsive_magnetic_PD}
\end{figure}

The magnetic polarizability $\alpha'_{\text{m}}$  of 162~nm dia. Si nanoparticles is positive yet smaller than its electric counterpart (refer to Fig.~\ref{fig:fig_SCM_data_matrix_optical_properties}b). $F_{\text{MD}}^{\text{PD}}$ and $F_{\text{ED}}^{\text{PD}}$ are calculated using Eqs.~\ref{eqn:PDA_e} and \ref{eqn:PDA_m} and shown in Fig.~\ref{fig:repulsive_magnetic_PD}c. The magnetic dipole force on the Si nanoparticle is toward the beam center. However, this is opposed by the electric dipole force that is directed outward and has a greater magnitude. Consequently, the particle is trapped within the annular region of the azimuthally polarized beam where the electric field amplitude is maximum. This conclusion agrees with the experimentally measured and ED-LD simulated dynamics of small-sized Si nanoparticles (see Fig.~\ref{fig:fig5}b in the main text).

Conversely, the 770~nm trapping laser is nearly maximally on-resonance with the magnetic dipole resonance of the 205~nm dia. Si nanoparticle as seen in the scattering spectrum shown in Fig.~\ref{fig:repulsive_magnetic_PD}a. This could cause a strong interaction with the longitudinal magnetic field $H_z$. Therefore, one might expect the magnetic dipole force to play the decisive role in the trapping behavior as compared to the electric counterpart. However, this expectation is not substantiated by the force spectra plotted in Fig.~\ref{fig:repulsive_magnetic_PD}c, \yl{where a lightly repulsive magnetic dipole force is found instead.} This unexpected behavior can be understood by inspecting the magnetic polarizability shown in Fig.~\ref{fig:repulsive_magnetic_PD}a. Near the magnetic dipole resonance, $\alpha'_{\text{m}}$ is close to but less than zero \yl{for a 205~nm dia. Si nanoparticle} resulting in the unexpected direction of the net force. \yl{A significant repulsive magnetic dipole force is found for the 239~nm Si nanoparticle since the polarizability is now strongly negative.} 

The calculated repulsive nature of the optical magnetic dipole force seems to be in contradiction with the strong trapping \yl{of larger Si nanoparticles at the center of the azimuthally polarized beam ( where $H_z$ is maximal) that is observed in the experiments. Hence, it is evident from the contradictory experimental analyses that another type of force must be responsible for the trapping behaviors of larger Si nanoparticles.}
\newpage

\subsection{Verification of the transverse scattering force and the scattering-corrected model (SCM)}
\label{SI_subsect:SCM}

\begin{figure}[H]
	\centering
	\hspace*{-3.5cm}
	\includegraphics[clip,trim=2.5cm 0cm 2.5cm 0cm,scale=0.8]{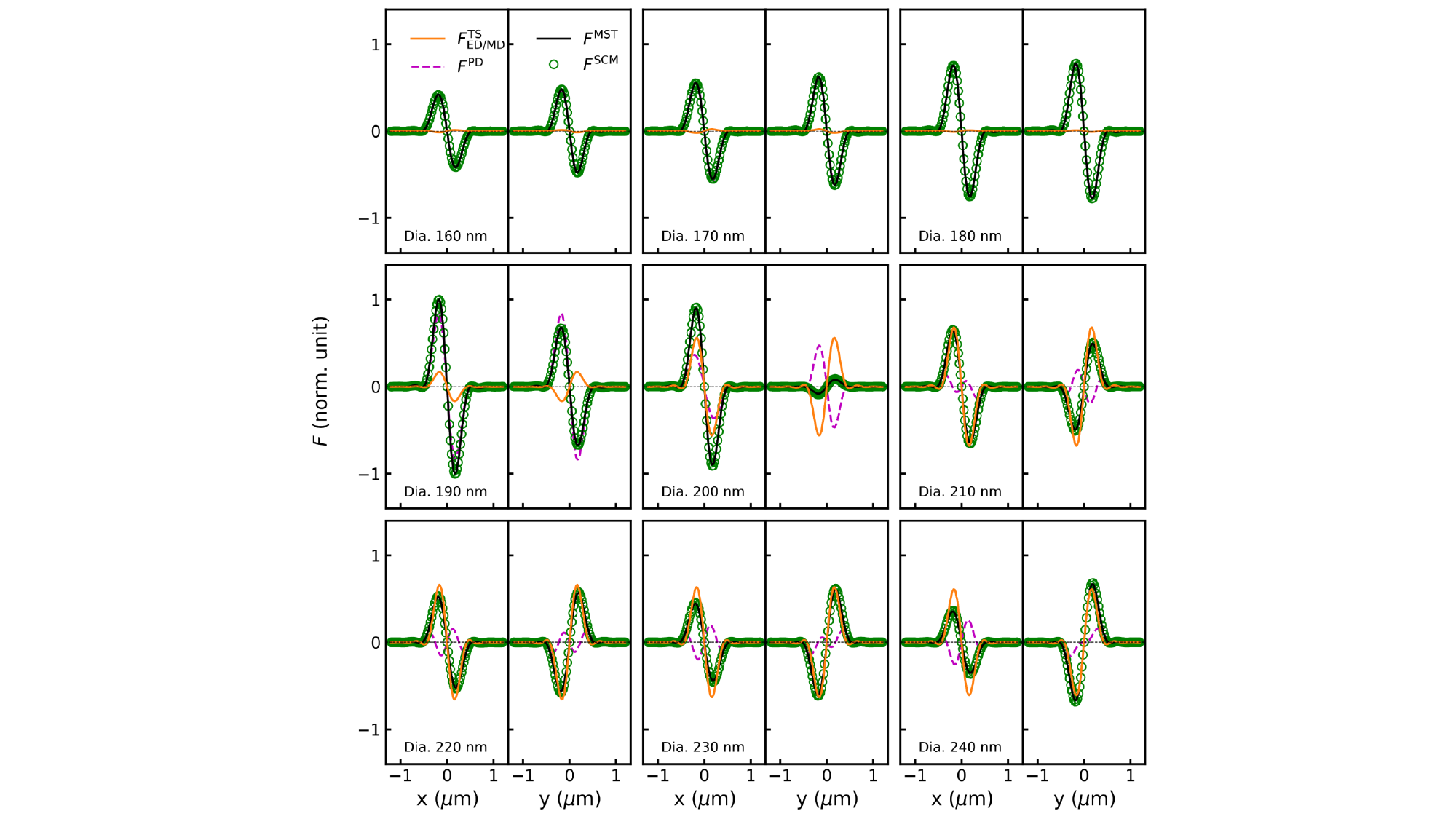}
	\caption{\textbf{Analysis of optical force on Si nanoparticles when trapped off-, near- and \yl{on-resonance with the optical magnetic dipole mode}.}
		Force analyses of Si nanoparticles with sizes ranging from 160~nm to 240~nm are performed using the same trapping configure, procedure, and method as in Fig.~3D. The horizontal axis is the displacement of the Si nanoparticle from the center of the Gaussian beam. 
	}
	\label{fig:fig_SCM_data_matrix}
\end{figure}

In order to probe the behavior of the dipole-governed transverse scattering (TS) force and validate our proposed SCM, force analyses are carried out for different size Si nanoparticles that span the trapping situations of interest: off-, near- and on-resonance with the optical magnetic dipole mode. These forces are evaluated with the same procedure and method used in Fig.~3D; hence, the data should be read in the identical manner. Noting that all forces are normalized by the maximum value of the MST-calculated forces for ease of comparison.

\begin{figure}[H]
	\hspace*{-0.5cm}
	\centering
	\hspace*{-0cm}
	\includegraphics[clip,trim=1.5cm 4.5cm 2cm 3.5cm,scale=0.6]{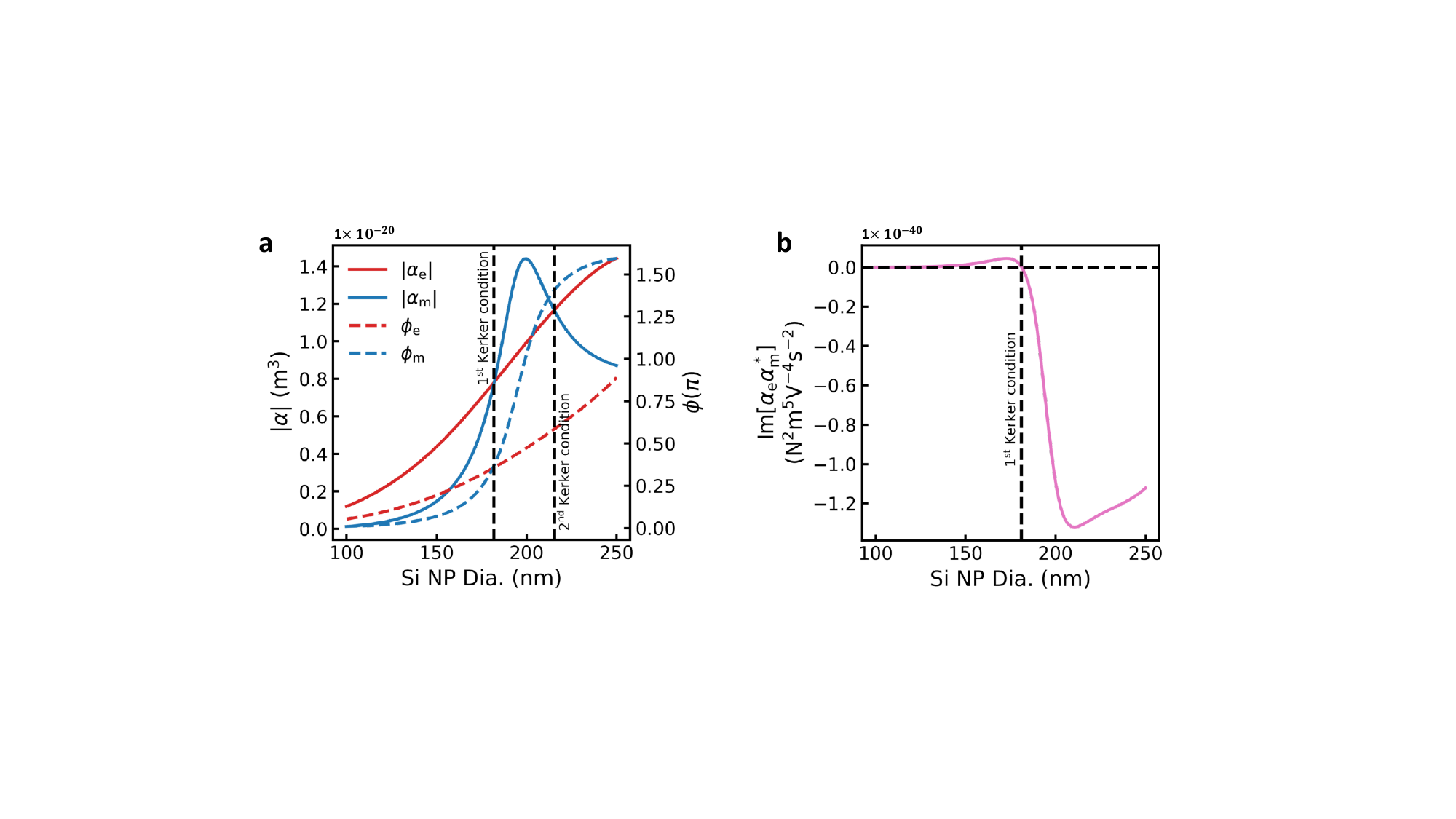}
	\caption{\textbf{\yl{Polarizability} properties of Si nanoparticles.}
		\textbf{(a)} Magnitudes ($|\alpha_{\text{e}}|$ and $|\alpha_{\text{m}}|$) and phases ($\phi_{\text{e}}$ and $\phi_{\text{m}}$) of the electric and magnetic polarizability of Si nanoparticles computed as a function of particle diameter. Two vertical dashed lines mark the dimensions that satisfy the first and second Kerker conditions. Note, the magnitudes are plotted in cgs units allowing direct comparison.
		\textbf{(b)} Imaginary part of the product of the electric and magnetic polarizabilities is plotted, with horizontal and vertical dashed lines highlighting the zero value and the first type Kerker condition, respectively. Note, SI units are used in this plot.
	}
	\label{fig:fig_SCM_data_matrix_optical_properties}
\end{figure}

Two remarkable conclusions can be drawn from the results shown in Fig.~\ref{fig:fig_SCM_data_matrix}, one of which is obvious while the other requires further scrutiny in conjunction with the parameters shown in Fig.~\ref{fig:fig_SCM_data_matrix_optical_properties}. The first is how well the SCM accounts for the electrodynamic forces that are absent from the point dipole model, particularly in \yl{the size region associated with on-resonance trapping}. The curves of both the $F^{\text{SCM}}$ and $F^{\text{MST}}$ are in perfect agreement for the entire range of particle sizes. When the particle size is small (Dia.$\le$180~nm), the scattering forces $F_{\text{ED/MD}}^{\text{TS}}$ are negligible, meaning that the point dipole model is accurate in capturing the optical forces. This is evident from the good agreement between the curves of $F^{\text{PD}}$ and $F^{\text{MST}}$. Increasing the particle size transitions the system into the on-resonance regime (for the 770~nm trapping laser) where the contributions from the transverse scattering force $F_{\text{ED/MD}}^{\text{TS}}$ becomes pronounced. As a result, the accuracy of the point dipole model deteriorates as seen in the growing discrepancy between $F^{\text{PD}}$ and $F^{\text{MST}}$. However, including $F_{\text{ED/MD}}^{\text{TS}}$ allows agreement with $F^{\text{MST}}$, thus affirming that the TS forces are the missing feature of the point dipole model. 

\yl{The second conclusion concerns the spatial anisotropy of the dipole-governed TS forces; that is, the scattering forces opposite signs along the orthogonal (x- and y-axis) directions for a Gaussian trapping beam linearly polarized along the x-axis (see Fig.~\ref{fig:num-ana_TSFs_Gaussian}). This finding holds even for the small Si nanoparticles. The root cause of this anisotropic phenomenon is discussed in Methods ``Analytical approximation of transverse scattering force'' in association with Eq~\ref{eqn:tsf-ana}.~ The interesting finding revealed from numerical calculations is the reversal of the scattering force that occurs between the small- (dia.$\le$180~nm) and large-sized (dia.$\ge$190~nm) Si nanoparticles. The small-sized Si nanoparticles experience a repulsive magnetic field-associated TS force ($F_{\text{MD}}^{\text{TS}}$) along the x-axis but an attractive electric field-actuated TS force ($F_{\text{ED}}^{\text{TS}}$) along the y-axis. The anisotropic behaviors of these two field-associated TS forces are reversed for the large Si nanoparticles. According to Eq.~\ref{eqn:tsf-ana} and shown in Fig.~\ref{fig:fig_SCM_data_matrix_optical_properties}b, this reversal is attributed to the Im$\left[\alpha_{\text{e}}\alpha_{\text{m}}^{*}\right]$ term, which changes sign from positive to negative at the size labeled as the first type Kerker condition. Based on the convention of the Kerker conditions\citesuppl{kerker1983electromagnetic}, and illustrated by the plots in Fig.~\ref{fig:fig_SCM_data_matrix_optical_properties}a, the first type Kerker condition requires the equality of both the magnitude and phase of the electric and magnetic polarizabilities, giving rise to an in-phase interference of the electric and magnetic dipolar fields. In contrast, the second Kerker condition occurs with a $\pi/2$ phase shift between the polarizabilities, causing the respective fields to manifest $\pi$ out-of-phase destructive interference. It is apparent from Fig.~\ref{fig:fig_SCM_data_matrix_optical_properties}a that only the first Kerker condition affects the TS forces' change of sign with size.    

Finally, we note that Fig.~\ref{fig:fig_SCM_data_matrix_optical_properties}b shows that the Im$\left[\alpha_{\text{e}}\alpha_{\text{m}}^{*}\right]$ term goes  to zero for small Si nanoparticles (dia.$<$130~nm). Therefore, the transverse scattering forces vanish for these small nanoparticles and conventional electric dipole trapping as described in the Mie-corrected point dipole approximation are obtained. therefore, the transverse scattering forces associated with the optical magnetic dipole resonance and described by Eq.~\ref{eqn:tsf-ana} can only be obtained for large Si nanoparticles.}

\newpage 

\subsection{Evaluation of the analytical approximation of the TS forces in both tightly and loosely focused Gaussian beams}
\label{SI_subsect:TSF_analytical}

\begin{figure}[H]
	\centering
        \hspace*{1.5cm}
	\includegraphics[clip,trim=6cm 3cm 0cm 3cm,scale=.8]{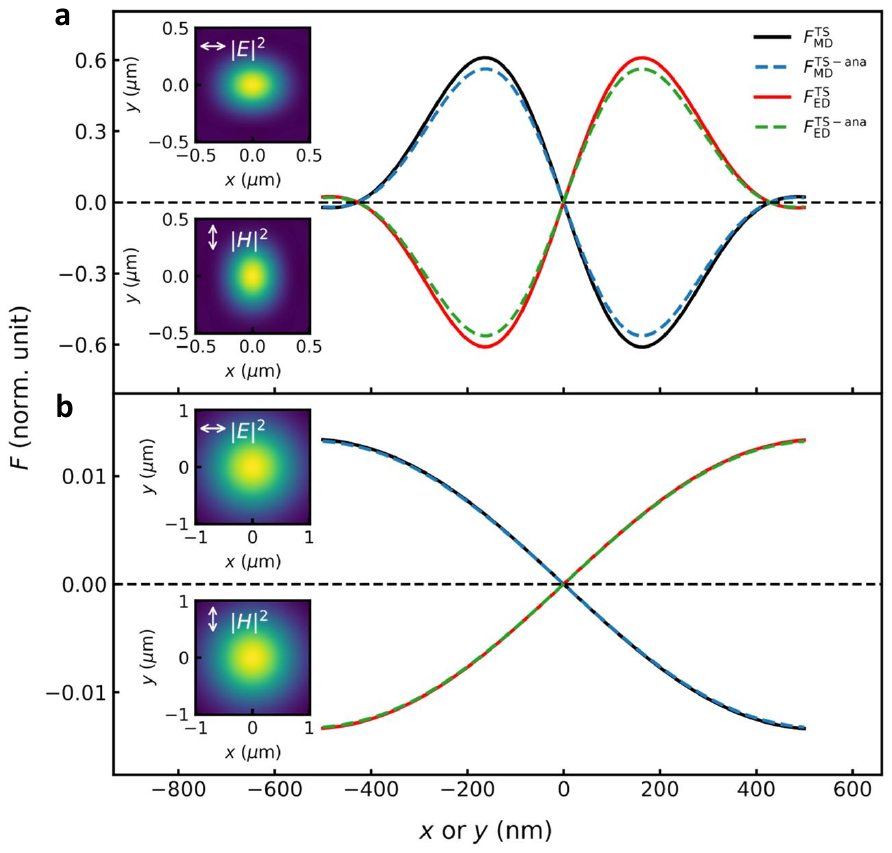}
	\caption{\textbf{Investigating the accuracy of the analytical approximation of dipole-governed TS forces.}
		\textbf{(a)} Numerically (solid) and analytically (dashed) calculated scattering forces acting on a 200~nm Si nanoparticle evaluated when scanning the particle along the x- ($F_{\text{MD}}^{\text{TS}}$: black, and $F_{\text{MD}}^{\text{TS-ana}}$: blue) and y-axis ($F_{\text{ED}}^{\text{TS}}$: red and $F_{\text{ED}}^{\text{TS-ana}}$: green) in a tightly focused Gaussian beam (beam width$\sim$300~nm). The energy densities of the respective fields and associated polarization directions are shown in the insets.
		\textbf{(b)} The comparison is repeated in a loosely focused Gaussian beam (beam width$\sim$1~\textmu m), as delineated by the images of the beam transverse profiles shown in the insets.
	}
	\label{fig:num-ana_TSFs_Gaussian}
\end{figure}

Figure~\ref{fig:num-ana_TSFs_Gaussian} shows the performance of the analytical approximation of the transverse scattering (TS) force (Eq.~\ref{eqn:tsf-ana}) in both tightly and loosely focused Gaussian beams with linear polarization (marked by the arrows in the insets) as assessed by comparing it with the numerical method (Eq.~\ref{eqn:TSF}). The high accuracy of the latter was validated in Sect.~\ref{SI_subsect:SCM}. When the Gaussian beam is loosely focused and the paraxial approximation is valid, the energy densities of the electric and magnetic fields shown in the insets of Fig.~\ref{fig:num-ana_TSFs_Gaussian}b are spatially indistinguishable, indicating that Eq.~\ref{eqn:tsf-ana-para} and \ref{eqn:tsf-ana} are equivalently same. The excellent agreement with the numerical result (of Eq.~\ref{eqn:TSF}) confirms the accuracy of the analytical approximation in a paraxial beam.

For a tightly focused Gaussian beam as shown in Fig.~\ref{fig:num-ana_TSFs_Gaussian}a, the energy densities of the electric and magnetic fields start to be spatially distinct from each other. The energy densities become elongated in orthogonal directions in relation to the respective electric and magnetic field polarization directions. The anisotropy in the spatial distributions indicates an inaccuracy of Eq.~\ref{eqn:tsf-ana-para}, which assumes uniform energy densities to obtain the TS forces. By contrast, Eq.~\ref{eqn:tsf-ana} accounts for the anisotropy of the fields, resulting in a satisfactory approximation relative to the numerically calculated TS forces. These comparisons affirm the discussion in Sect.~\ref{SI_subsect:SCM} that the applicability of the analytical approximation for a tightly focused beam, and hence its utility in unraveling the nature of the dipole-governed TS forces.

\newpage

\subsection{Decomposition of on-resonance TS forces in an azimuthally polarized beam}
\label{SI_subsect:Decomp_on-res_TSF}

\begin{figure}[H]
	\centering
        \hspace*{-3.7cm}
	\includegraphics[clip,trim=2cm 4cm 2cm 3cm,scale=.8]{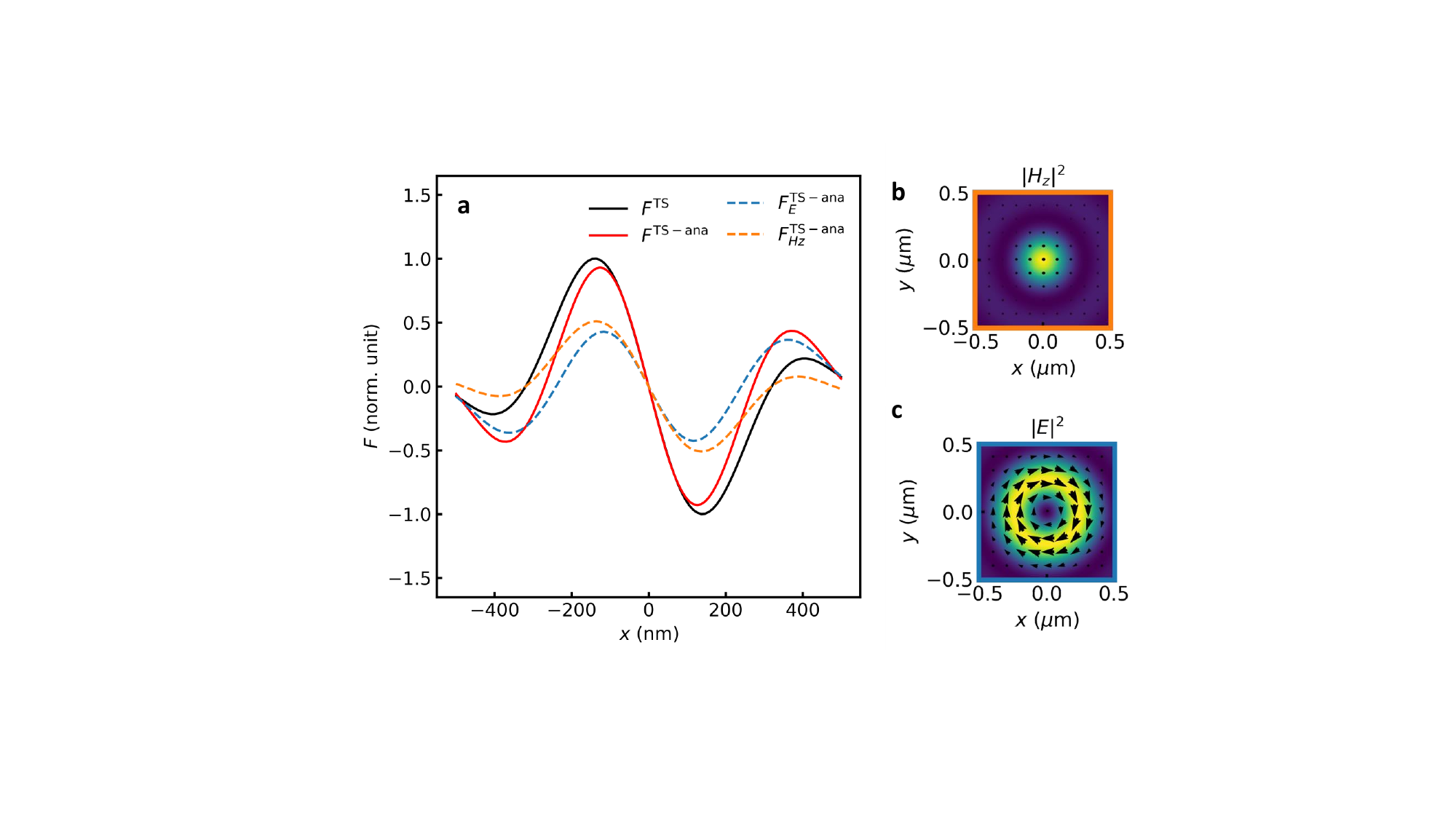}
	\caption{\textbf{Decomposition of the TS force acting on a 200~nm dia. Si nanoparticle trapped in an azimuthally polarized beam.}
		The TS force is calculated numerically (black) using Eq.~\ref{eqn:TSF} and analytically (red) using Eq.~\ref{eqn:tsf-ana}, where the analytical result is the sum of the contributions from electric ($E$ shown in \textbf{(c)}) and magnetic ($H_z$ shown in \textbf{(b)}) fields. The respective fields are shown, with black arrows indicating the polarization directions. Note, the black dots in the rendering of the magnetic field $H_z$ demonstrate the field's longitudinal polarization along the z-axis.
	}
	\label{fig:tsf_decomposition}
\end{figure}

\yl{Figure~\ref{fig:tsf_decomposition} shows the TS forces acting on a 200~nm Si nanoparticle obtained from simulations in a tightly focused azimuthally polarized beam. The TS forces experienced by the Si nanoparticles are for the case of on-resonance trapping. Equation~\ref{eqn:TSF} (solid black curve) is used to accurately determine the TS force acting on the particle. The analytical approximation (Eq.~\ref{eqn:tsf-ana}; solid red curve) is used to determine the TS force's constituents attributed to the azimuthally polarized electric $E$ field in the annular region and the longitudinally polarized magnetic field $H_z$ at the center as shown in Fig.~\ref{fig:tsf_decomposition}c and b, respectively. Comparing the results of these two methods, their similarity, particularly in the central region of the beam, affirms the validity of the analytical approximation (Eq.~\ref{eqn:tsf-ana}) as well as its representation of the electric and magnetic field contributions. As seen in Fig.~\ref{fig:tsf_decomposition}, the $H_z$ associated TS force ($F_{H_z}^{\text{TS-ana}}$) drives the 200~nm dia. Si nanoparticle toward the magnetic field maximum, while the $E$ associated TS force ($F_{E}^{\text{TS-ana}}$) simultaneously repels the particle away from annular region where the electric field is maximized. This repulsion causes the confinement of the particle in the central region of the beam. In the case of the 200~nm dia. Si nanoparticle, the contribution from the radially polarized magnetic field $H_{\rho}$ along the radial direction is negligibly small, in part because the magnitude of the $H_{\rho}$ is relatively small in the tightly focused beam (refer to Fig.~\ref{fig:fig1}b in the main text). Therefore, The TS force actuated by $H_{\rho}$ is not important in the present work.}

\newpage

\subsection{Analysis of optical forces exerted on the 205~nm dia. Si nanoparticle}
\label{SI_subsect:anaF_205nmSiNP}

\begin{figure}[H]
	\hspace*{1.5cm}
	\centering
	\includegraphics[clip,trim=6.5cm 4cm 0cm 4cm,scale=.9]{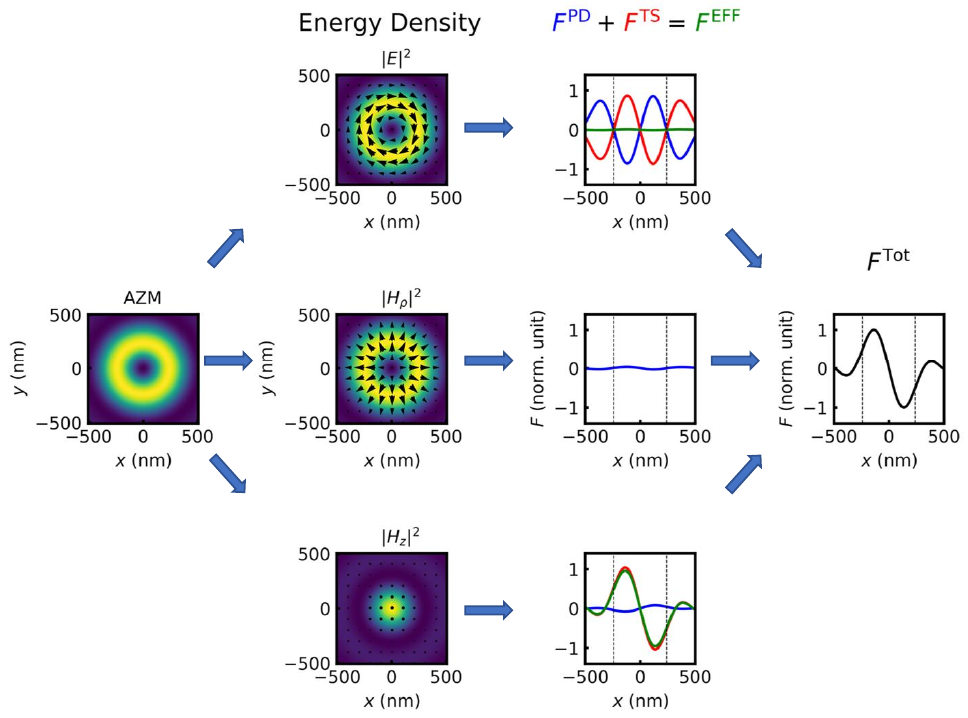}
	\caption{\textbf{Demonstration of the optical forces experienced by the 205~nm dia. Si nanoparticle trapped in a tightly focused azimuthally polarized beam (1st column).}
		(2nd column) The beam is decomposed into three constituent fields: azimuthally polarized electric field $E$, radially polarized magnetic field $H_{\rho}$, and longitudinally polarized induced magnetic field $H_z$. The arrows displayed in the energy density profiles indicate the polarization of the respective fields. (3rd column) Each field creates an optical force ($F^{\text{EFF}}$, green) consisting of a dipole force ($F^{\text{PD}}$, blue) and/or a scattering force ($F^{\text{TS}}$, red). The vertical dotted lines in the force plots indicate the positions of the location of the maximum values of the electric field annulus. (4th column) Combining these separate contributions gives rise to the total optical force $F^{\text{Tot}}$ that determines the particle's dynamics observed in both experiment and ED-LD simulation.
	}
	\label{fig:force_analysis_205sinp}
\end{figure}

Figure~\ref{fig:force_analysis_205sinp} shows the optical force ($F^{\text{Tot}}$) exerted on the 205~nm dia. Si nanoparticle trapped in the azimuthally polarized beam. The total force is decomposed into individual components delineated by the constituent fields (i.e., $E$, $H_{\rho}$, and $H_z$) of the beam and associated categories of forces (i.e., dipole $F^{\text{PD}}$ and scattering $F^{\text{TS}}$ forces). The scattering forces are evaluated with the analytical method (Eq.~\ref{eqn:tsf-ana}) that was quantitatively examined and validated in Sect.~\ref{SI_subsect:Decomp_on-res_TSF}. 

According to the force plots of Fig.~\ref{fig:force_analysis_205sinp}, the electric field $E$ produces a balanced situation where the dipole and scattering forces are nearly equal in strength but opposite in direction, causing their combined trivial influence on the particle's dynamics. Likewise, the impact of the radially polarized magnetic field $H_{\rho}$ on the particle's trapping behavior is minimal, as reflected in its very small dipole force. This $H_{\rho}$-associated force is very small because of the near-zero magnetic polarizability of particles of this size (see Fig.~\ref{fig:repulsive_magnetic_PD}a). In fact, the force is repulsive (from the beam center) due to the negative magnetic polarizability of the large 205~nm dia. particle size and associated resonance peak of the magnetic dipole mode. On its own, this $H_{\rho}$-associated force would push the particle away from the annular region with the $H_{\rho}$ towards the central region. The associated scattering force has intentionally been excluded from consideration, mainly because this force has no contribution along the radial direction, as explained in Sect.~\ref{SI_subsect:Decomp_on-res_TSF}.

In contrast with the above-mentioned two fields, the longitudinal magnetic field $H_z$ exerts the decisive influence on the electrodynamics of the 205~nm dia. nanoparticle, as evident from its associated scattering force that is dominant over other forces and pushes the particle toward the central region where the induced magnetic field is maximum. In addition, the longitudinal magnetic field $H_z$ also exerts a repulsive dipole force on the particle because of the negative magnetic polarizability $\alpha_{\text{m}}^{'}$ of the Si nanoparticle (see Sect.~\ref{SI_subsect:Repulsive_MDF}). However, the magnitude of $\alpha_{\text{m}}^{'}$ is close to zero, leading to the repulsive dipole force that is sufficiently small to be negligible.

\newpage

\subsection{Analysis of optical forces exerted on the 239~nm dia. Si nanoparticle}
\label{SI_subsect:anaF_239nmSiNP}

\begin{figure}[H]
	\hspace*{1.5cm}
	\centering
	\includegraphics[clip,trim=6.5cm 4cm 0cm 4cm,scale=.9]{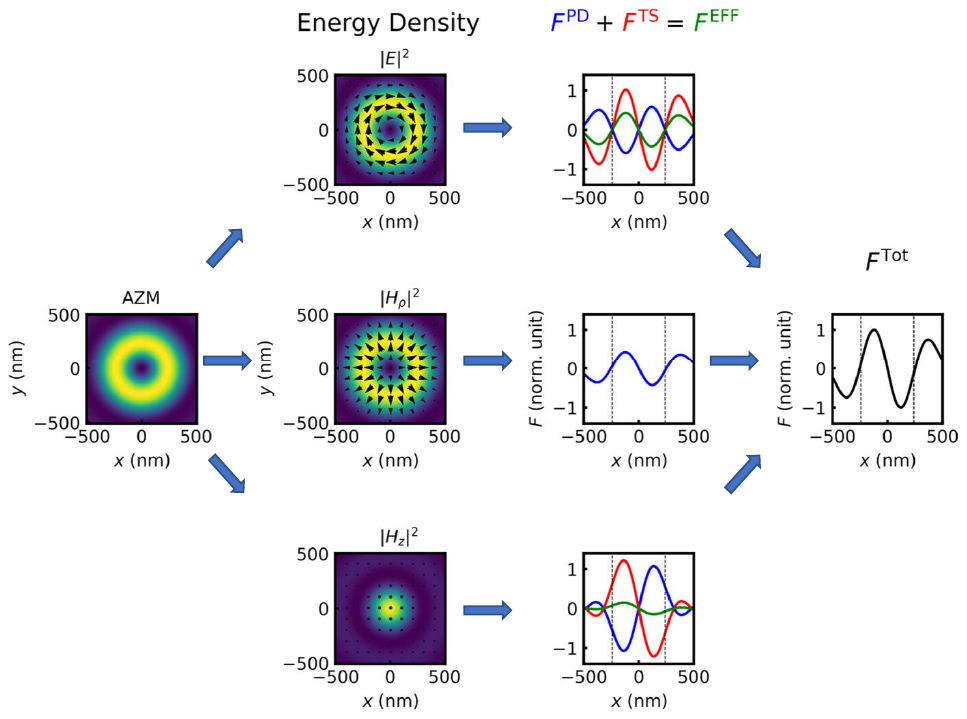}
	\caption{\textbf{Demonstration of the optical forces experienced by the 239~nm dia. Si nanoparticle trapped in the tightly focused azimuthaly polarized beam.}
		The figure follows the same format as Fig.~\ref{fig:force_analysis_205sinp}. (2nd column) The beam is decomposed into three constituent fields: azimuthally polarized electric field $E$, radially polarized magnetic field $H_{\rho}$, and longitudinally polarized induced magnetic field $H_z$. The arrows displayed in the energy density profiles indicate the polarization of the respective fields. (3rd column) Each field creates an optical force ($F^{\text{EFF}}$, green) consisting of a dipole force ($F^{\text{PD}}$, blue) and/or a scattering force ($F^{\text{TS}}$, red). The vertical dotted lines in the force plots indicate the positions of the location of the maximum values of the electric field annulus. (4th column) Combining these separate contributions gives rise to the total optical force $F^{\text{Tot}}$ that determines the particle's dynamics observed in both experiment and ED-LD simulation.}
	\label{fig:force_analysis_239sinp}
\end{figure}

Figure~\ref{fig:force_analysis_239sinp} shows the optical force and its decomposition for a large 239~nm dia. Si nanoparticle. Compared to the 205~nm Si nanoparticle, the optical force associated with the electric field $E$ becomes important in addition to the optical force associated with and magnetic field $H_{z}$. Given the electric and magnetic polarizability of the particle (see Fig.~\ref{fig:repulsive_magnetic_PD}a and Fig.~\ref{fig:fig_SCM_data_matrix}b and c), the scattering and dipole forces associated with the $H_z$ are closely matched in strength (see lower panel in Fig.~\ref{fig:force_analysis_239sinp}), which results in a cancellation that essentially eliminates the field's impact on the particle's dynamics. By contrast, the forces resulting from the $E$ and $H_{\rho}$ (see upper panel in Fig.~\ref{fig:force_analysis_239sinp}) become the primary factors that constrain the particle in the central region through repulsive forces. However, the underlying mechanisms behind the $E$- and $H_{\rho}$-associated forces are different. The repulsion due to $H_{\rho}$ originates from the magnetic dipole, while the repulsion due to the $E$-field stems from transverse scattering force. The origin of the repulsive magnetic dipole force has been discussed in Fig.~\ref{fig:force_analysis_205sinp}. As for the $E$-associated scattering force, its rapid growth is attributed to the particle becoming on-resonance with the electric dipole mode, which significantly increase the electric dipole radiation. Thus, in addition to the optical forces mediated by the magnetic fields, the scattering force associated with the electric field acts as an additional contribution, enhancing the effectiveness of the azimuthally polarized beam in confining the particle to the center of the beam.

\newpage

\subsection{Defining photonic Hall effect {PHE}}
\label{SI_subsect:PHE}

\begin{figure}[H]
	\vspace*{-1.8cm}
	\centering
	\includegraphics[clip,trim=6cm 3.5cm 6cm 0cm,scale=.8]{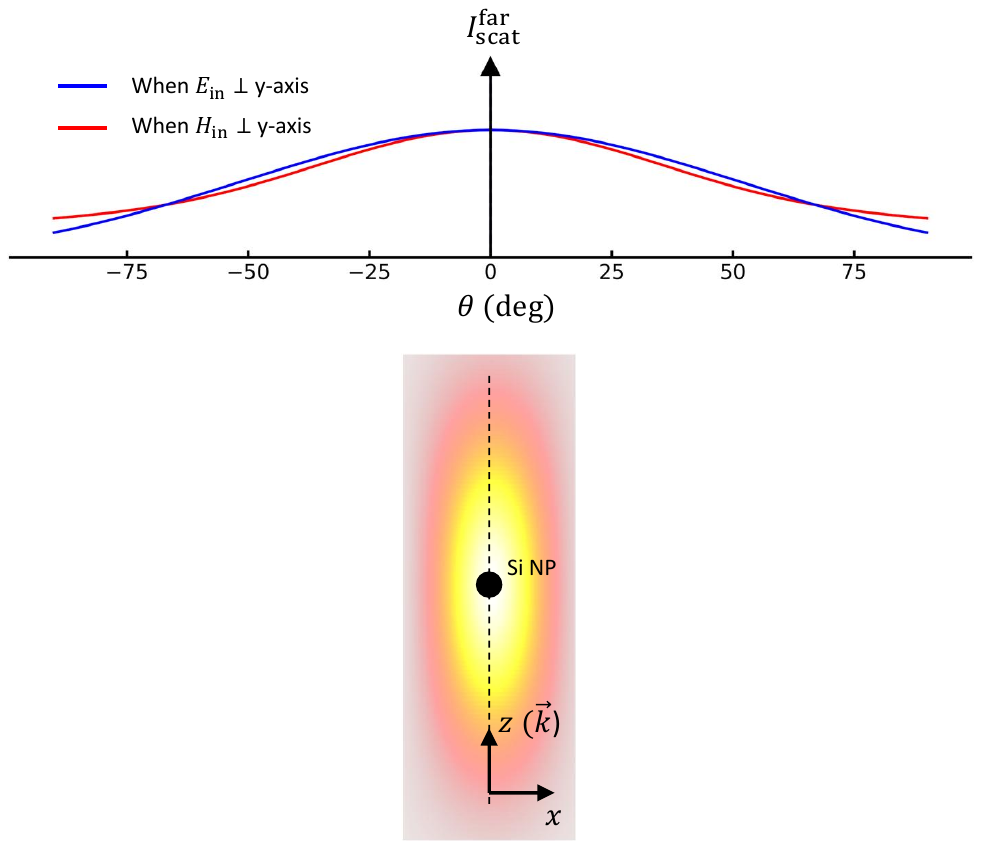}
	\caption{\textbf{Polarization-dependent scattering from a single Si nanoparticle placed at the center of a Gaussian beam.}
    	A 205~nm dia. Si nanoparticle centered at the focus of a linearly polarized Gaussian beam is simulated to characterize its far-field scattering behaviors in response to the two orthogonal polarization states of the excitation light: one with its electric field $E_{\text{in}}$ polarized along the y-axis (blue curve) and the other with its magnetic field $H_{\text{in}}$ polarized along the y-axis (red curve). The majority of the particle's far-field scattering, $I_{\text{scat}}^{\text{far}}$, is along the z-axis, that is for $\theta$=0 for both polarization states.
	}
	\label{fig:PHE_centered}
\end{figure}

Since its inception in electronic systems, the Hall effect has been routinely used to describe transport of charge carriers driven by applied electric fields in conductors/semiconductors or photons propagating through media with refractive index variations. Light has been proven to exhibit spin Hall effect, known as spin Hall effect of light\citesuppl{onoda2004hall,hosten2008observation}. This effect occurs when light crosses a planar or structured optical interface, leading to a transverse splitting of light into components carrying opposite spin angular momentum (i.e., right- and left-handed circular polarizations) because of spin-orbit interactions\citesuppl{onoda2004hall,hosten2008observation}. This principle extends to nanoparticles. 
Probing the energy flux in proximity to a nanoparticle illuminated by a beam reveals that the particle modulates its near-fields with opposite directions that depend on the incident beam's circular polarization (spin) states\citesuppl{haefner2009spin}. In addition, recent work has shown the possibility of enhancing the photonic spin Hall effect by breaking the parity symmetry of the systems\citesuppl{rodriguez2010optical,li2019photonic,neugebauer2018weak}. For instance, parity-symmetry breaking, induced by either the anisotropic geometry or spatial displacement of single nanoparticles from the beam center, allows dramatically altering the directionality of the particles' far-field scattering light by switching between the excitation beam's two opposite spin states\citesuppl{li2019photonic,neugebauer2018weak}.

\begin{figure}[H]
	\hspace*{-1cm}
	\centering
	\includegraphics[clip,trim=6cm 3.5cm 6cm 2cm,scale=.8]{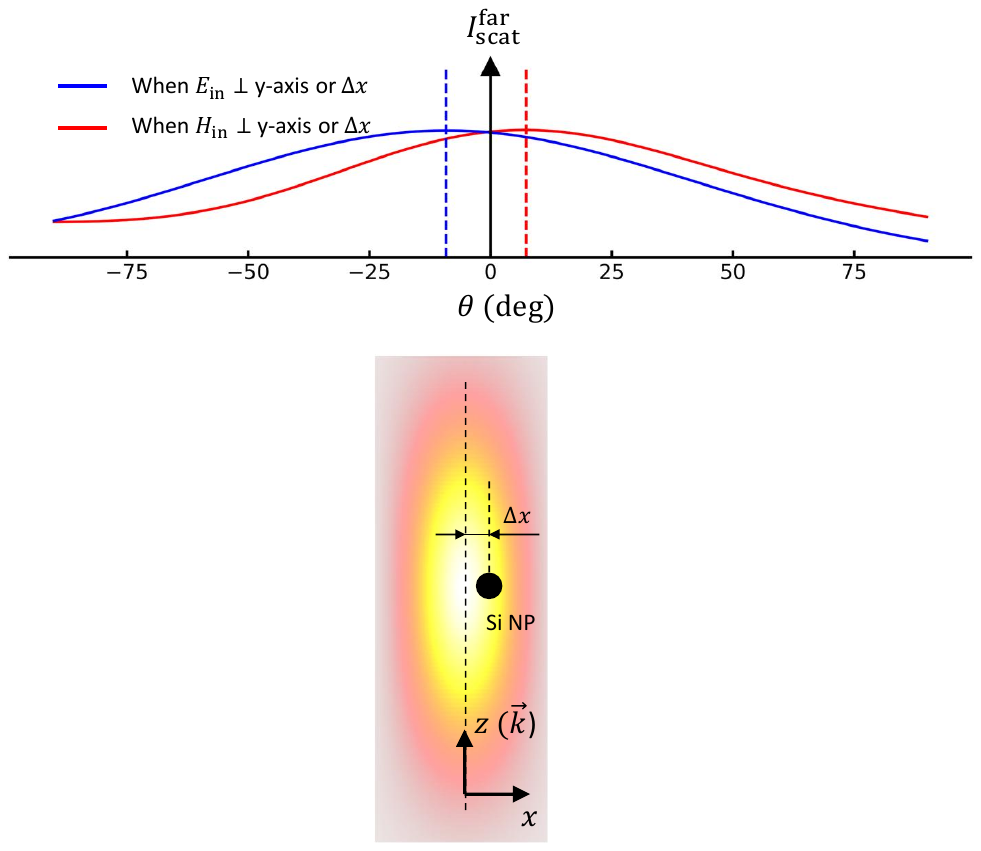}
	\caption{\yl{\textbf{Polarization-dependent scattering from a single Si nanoparticle displaced from the center of a Gaussian beam.}
		The figure depicts the simulation results in the identical style as Fig.~\ref{fig:PHE_centered}, with the distinction that the 205 nm diameter Si nanoparticle is displaced by half the beam width, $\Delta x$, along the x-axis. This horizontal displacement shifts the center of the angular distribution of the scattered light away from the z-axis in (opposite) directions that depend on the polarization state of the excitation light. The mean values of these angular distributions are indicated by the blue (for $E_{\text{in}}\perp\Delta x$) and red (for $H_{\text{in}}\perp\Delta x$) dashed lines, respectively.
	}}
	\label{fig:PHE_displaced}
\end{figure}

Our definition of the photonic Hall effect (PHE)  builds upon this previous work. In the present study, we identified a linear polarization-dependent directionality of far-field electromagnetic radiation scattered by single Si nanoparticles for an illumination condition that breaks parity-symmetry (mirror symmetry). Since this far-field scattering is associated with the excitation light's linear polarization state rather than its spin state, we refer to this phenomenon as the ``photonic Hall effect'' to differentiate it from its spin counterpart. How this effect take places is demonstrated in Figs.~\ref{fig:PHE_centered} and~\ref{fig:PHE_displaced}, where the far-field of scattering  light from a 205~nm dia. Si nanoparticle is calculated for two illumination conditions: (1) with the nanoparticle centered transversely in the Gaussian beam as shown in Fig.~\ref{fig:PHE_centered}; and (2) with the Si nanoparticle shifted away from the center along the x-axis by a distance $\Delta x$ as shown in Fig.~\ref{fig:PHE_displaced}. Note that the calculations are based on the GLMT as detailed in the Methods section.

Figure~\ref{fig:PHE_centered} shows that when the Si nanoparticle is positioned precisely at the center of a symmetric focused Gaussian beam, its far-field scattering intensity distribution, $I_{\text{scat}}^{\text{far}}$, is computed as a function of the polar angle, $\theta$. We consider the scattered intensity in the x-z plane and $\theta$ is relative to the z-axis (or incident light propagation direction, $\Vec{k}$). The plot illustrates that the particle predominantly scatters light along and symmetrically about the z-axis. This behavior persists when switching between the two orthogonal linear polarization states of the incident light: (1) the incident electric field $E_{\text{in}}\perp$ y-axis and (2) the incident magnetic field $H_{\text{in}}\perp$ y-axis (equivalently, $E_{\text{in}}\parallel$ y-axis). However, as shown in Fig.~\ref{fig:PHE_displaced}, when the particle is displaced from the center by a small distance, $\Delta x$, the angular distribution of the scattering is no longer independent of the linear polarization state of the incident beam. The displacement-induced (parity) symmetry breaking leads to a shift of the peak (and mean) of the angular scattering distributions (as indicated by the dashed vertical lines) in opposite directions relative to the z-axis. Each distribution, with opposite angular shifts, corresponds to one linear polarization state of the incident light. This dependence of the scattering directionality on the incident light's linear polarization state manifests the proposed photonic Hall effect. This PHE results in what we have termed "transverse scattering forces," which, according to Eqs.~\ref{eqn:tsf} and \ref{eqn:tsf-ana}, are linked to the exciting electric and magnetic fields and demonstrate their quasi-anisotropy. These characteristics have not been discovered by any previous studies concerning the optical trapping behaviors influenced by interactions between light's magnetic component with matter\citesuppl{chen2011optical,duan2022asymmetric,duan2021transverse}.

\newpage

\subsection{\yl{Evaluation of the suitability other materials for magnetic field-mediated optical trapping}}
\label{SI_subsect:variousMaterials}

\begin{figure}[H]
	\hspace*{-0cm}
	\centering
	\includegraphics[clip,trim=0cm 0cm 0cm 0cm,scale=.45, angle=0] {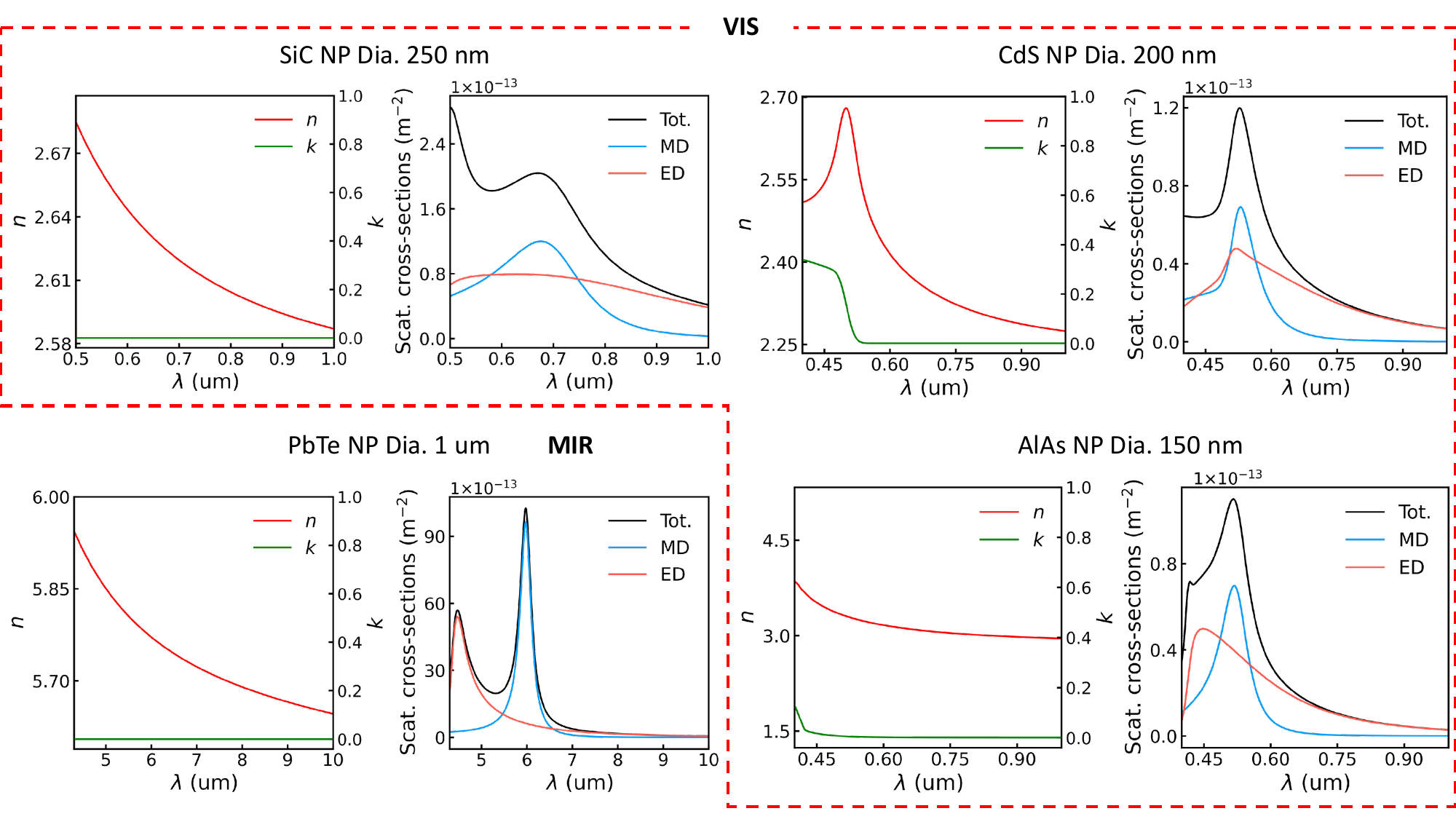}
	\caption{\textbf{Calculated scattering cross-section of four nanoparticle materials that support MD resonances.}
		These materials, SiC, CdS, PbTe, and AlAs, are chosen for their large refractive indices ($n$) and small extinction coefficients ($\kappa$), highlighting their advantages in field confinement and minimal absorption, respectively. When these materials are formed into particles of suitable sizes their ED (red curve) and MD (blue curve) modes become spectrally distinct. This distinction is a promising indication of their suitability for optical trapping influenced by magnetic fields (as illustrated in Fig.~\ref{fig:varMaterials_trapping}). The materials are grouped for their suitability for operation across different wavelength regimes: the visible regime (SiC, CdS, and AlAs in the red dashed enclosure) and the mid-infrared regime (PbTe).  All simulation results assume an aqueous (n=1.33) environment.
	}
	\label{fig:varMaterials}
\end{figure}

\yl{In this section, we show that the magnetic field-mediated optical trapping and its associated transverse scattering optical forces that are demonstrated in this paper for Si nanoparticle is applicable to a wide range of materials. These nanoparticles and materials may have other useful functionalities. Furthermore, nanoparticles made of other materials are candidates for nano-/macro-particle manipulations across the visible and mid-infrared spectral regions. To demonstrate this, Figure~\ref{fig:varMaterials} shows the scattering cross-sections for nanoparticles made of four commonly used semiconductors (i.e., SiC\citesuppl{singh1971nonlinear}, CdS\citesuppl{gomez1985weighted}, PbTe\citesuppl{weiting1990temperature}, and AlAs\citesuppl{fern1971refractive},
) characterized by high refractive indices, $n$, and low extinction coefficients, $\kappa$, for the effective field confinement and suppression of thermal effects. Analogous to Si, nanoparticles of these materials exhibit modified optical Mie scattering spectra (black curves) featured by the ED (red curves) and MD (blue curves) components due to their individual complex indices of refraction and boundary conditions. With appropriately engineered sizes for each material, they all exhibit pronounced MD modes, with the resonance peaks spectrally separated from those of the ED modes, implying their suitability for magnetic field-mediated optical trapping.  The efficacy of the MD modes can be further enhanced if the ambient condition is switched to air or vacuum (n=1.0) due to the increased refractive index contrast. Figure~\ref{fig:varMaterialsAir} shows the Mie scattering cross-section of a SiC nanoparticle in air, where its MD's increases as the resonance narrows giving a higher Q-factor compared to its MD mode when the same size SiC nanoparticle is in aqueous solution (shown in Fig.~\ref{fig:varMaterials}).

\begin{figure}[t]
	\hspace*{-0.5cm}
	\centering
	\includegraphics[clip,trim=3cm 2cm 5cm 0cm,scale=.6, angle=0]{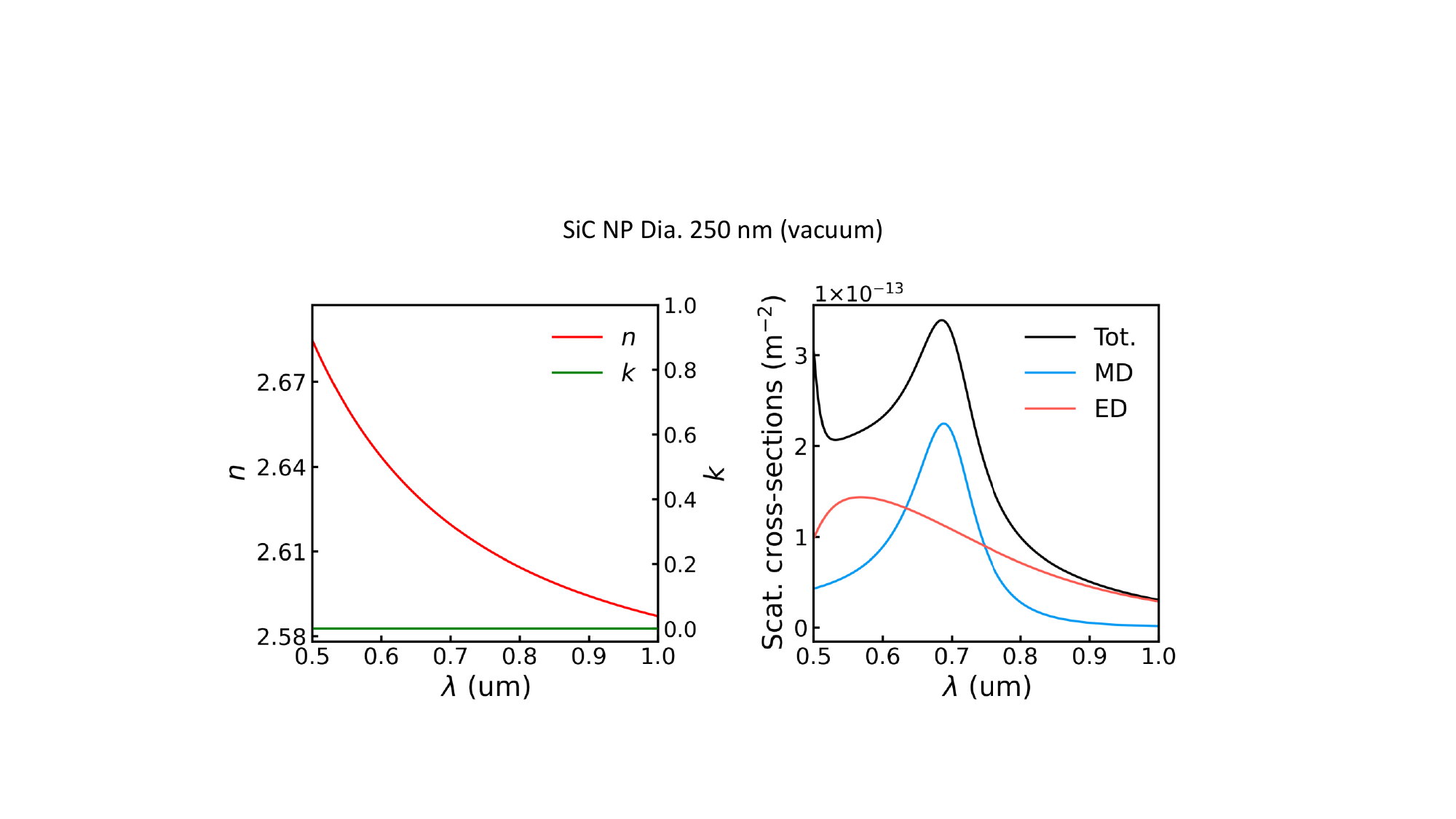}
	\caption{\yl{\textbf{The enhancement of the spectral distinction between the ED and MD modes under vacuum conditions.}
		The SiC nanoparticle case is highlighted here to show how an increased refractive index contrast -- specifically, changing from an aqueous solution (n=1.33 in Fig.~\ref{fig:varMaterials}) to vacuum (n=1.0) environment-- can enhance the spectral discrimination between the ED and MD modes. This enhancement suggests that magnetic field-mediated optical trapping is an effective technique for trapping these proposed nanoparticle materials in vacuum conditions.
	}}
	\label{fig:varMaterialsAir}
\end{figure}

\begin{figure}[H]
	\hspace*{-0.5cm}
	\centering
	\includegraphics[clip,trim=3cm 4cm 3cm 0cm,scale=.6]{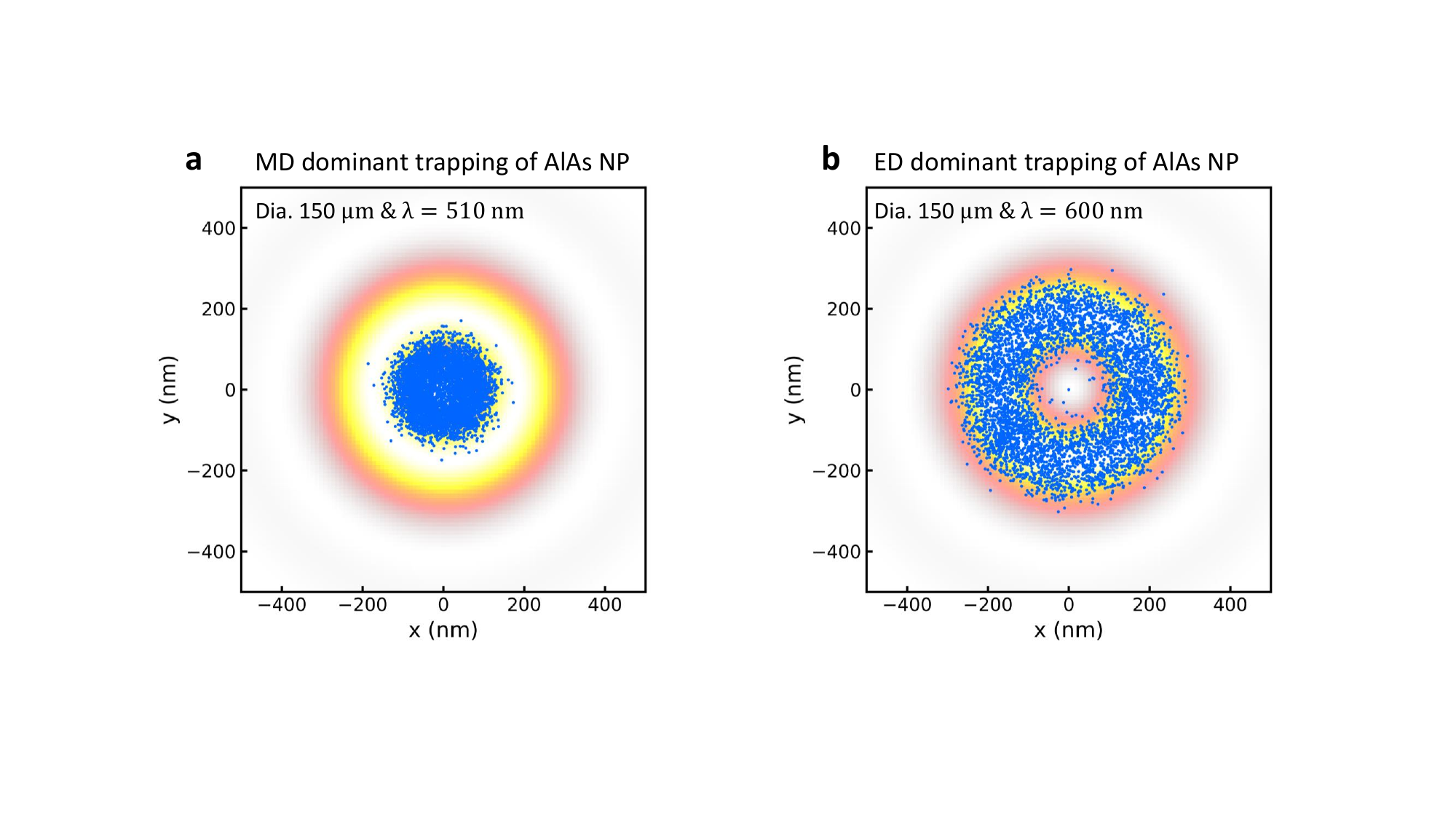}
	\caption{\yl{\textbf{ED-LD simulations of a 150~nm dia. AlAs nanoparticle trapped in an azimuthally polarized beam.}
		\textbf{(a)} Referring to the wavelength-dependent scattering cross-section of the AlAs nanoparticles shown in Fig.~\ref{fig:varMaterials}, setting the trapping laser wavelength to 510~nm corresponds to the peak of the MD mode. The ED-LD simulations (blue points) show the particle is primarily trapped by the longitudinally polarized magnetic field in the central area of the azimuthally polarized beam. \textbf{(b)} When the trapping laser's wavelength is detuned toward longer wavelengths, where the ED mode's magnitude becomes dominant (e.g., at 600 nm), the AlAs nanoparticle is mainly trapped in the annular region where it interacts more intensely with the electric field of light. Note, the properties of the trapping beam simulated here are the same as those described in Fig.~1 of the main text.
	}}
	\label{fig:varMaterials_trapping}
\end{figure}

Magnetic field-mediated optical trapping is demonstrated by the ED-LD simulations for AlAs shown in Fig.~\ref{fig:varMaterials_trapping}. This example illustrates the dynamics of the AlAs nanoparticle within an azimuthally polarized beam. The two distinct laser wavelength excitations underscore the two types of trapping mechanisms that are predominantly driven by the magnetic (left panel) and electric components (right panel) of light, respectively. Specifically, When the trapping laser wavelength is set at 510~nm, strongly on-resonance with the MD mode (see Fig.~\ref{fig:varMaterials}), the AlAs nanoparticle is trapped in the central region of the azimuthally polarized beam, a region dominantly occupied by the longitudinally polarized magnetic field $H_{z}$ (refer to Fig.~1 in the main text). Tuning the trapping laser to 600~nm makes the particle's ED mode dominant, which causes a shift of the trapping from the beam's center to its annular region due to overwhelming electric field-matter interactions. The trapping dynamics shown in Fig.~
\ref{fig:varMaterials_trapping}, which are consistent with those observed experimentally for Si nanoparticles, also apply to the other three materials (SiC, CdS, and PbTe). These simulation results affirm the broad applicability of optical magnetic field associated trapping.}

\renewcommand{\theenumiv}{\arabic{enumiv}}
\setcounter{enumiv}{0} 
\bibliographystylesuppl{naturemag}
\bibliographysuppl{scibib}